\documentclass[preprint]{aastex}

\usepackage{natbib}
\newcommand{\ec}{$\eta$\,Crv} \newcommand{\mic}{$\mu$m}

\shorttitle{Imaging of the $\eta$\,Crv debris disk with {\it Herschel}}
\shortauthors{Duch\^ene et al.}

\begin{document}

\title{Spatially resolved imaging of the two-component $\eta$\,Crv debris disk with {\it Herschel}}

\author{G. Duch\^ene\altaffilmark{1}, P. Arriaga} \affil{Astronomy Department,
  University of California, Berkeley, CA 94720, USA}

\and

\author{M. Wyatt, G. Kennedy} \affil{Institute of Astronomy, University of Cambridge, Madingley Road, Cambridge, CB3 0HA, UK}

\and

\author{B. Sibthorpe} \affil{SRON Netherlands Institute for Space Research, P.O. Box 800, 9700 AV Groningen, the Netherlands}

\and

\author{C. Lisse} \affil{JHU-APL, 11100 Johns Hopkins Road, Laurel, MD 20723, USA}

\and

\author{W. Holland\altaffilmark{2}} \affil{UK Astronomy Technology Centre, Royal Observatory Edinburgh, Blackford Hill, Edinburgh EH9 3HJ, UK}

\and

\author{J. Wisniewski} \affil{H.L. Dodge Department of Physics \& Astronomy, University of Oklahoma, 440 W Brooks St Norman, OK 73019, USA}

\and

\author{M. Clampin} \affil{NASA Goddard Space Flight Center, Code 681, Greenbelt, MD 20771, USA}

\and

\author{P. Kalas}\affil{Astronomy Department,
  University of California, Berkeley, CA 94720, USA}

\and

\author{C. Pinte\altaffilmark{1,3}} \affil{UMI-FCA, CNRS/INSU France (UMI 3386)}

\and

\author{D. Wilner} \affil{Harvard-Smithsonian Center for Astrophysics, 60 Garden Street, Cambridge, MA 02138, USA}

\and

\author{M. Booth} \affil{Instituto de Astrof\'isica, Pontificia Universidad Cat\'olica de Chile, Vicu\~na Mackenna 4860, 7820436 Macul, Santiago, Chile}

\and

\author{J. Horner\altaffilmark{4}} \affil{School of Physics, University of New South Wales, Sydney, NSW 2052, Australia}

\and

\author{B. Matthews\altaffilmark{5}}\affil{National Research Council of Canada Herzberg Astronomy \& Astrophysics, 5071 West Saanich Road, Victoria, BC V9E 2E7, Canada}

\and

\author{J. Greaves}\affil{SUPA, School of Physics and Astronomy, University of St. Andrews, North Haugh, St. Andrews KY16 9SS, UK}

\altaffiltext{1}{UJF-Grenoble 1 / CNRS-INSU, Institut de Plan\'etologie et d'Astrophysique (IPAG) UMR 5274, 38041 Grenoble, France}
\altaffiltext{2}{Institute for Astronomy, University of Edinburgh, Royal Observatory Edinburgh, Blackford Hill, Edinburgh EH9 3HJ, UK}
\altaffiltext{3}{Departamento de Astronom\'ia, Universidad de Chile, Santiago, Chile}
\altaffiltext{4}{Australian Centre for Astrobiology, University of New South Wales, Sydney, NSW 2052, Australia}
\altaffiltext{5}{Department of Physics and Astronomy, University of Victoria, Victoria, BC, V8W 3P6, Canada}

\begin{abstract}
We present far-infrared and sub-millimeter images of the \ec\ debris disk system obtained with {\it Herschel} and SCUBA-2, as well as {\it Hubble Space Telescope} visible and near-infrared coronagraphic images. In the 70\,\mic\ {\it Herschel} image, we clearly separate the thermal emission from the warm and cold belts in the system, find no evidence for a putative dust population located between them, and precisely determine the geometry of the outer belt. We also find marginal evidence for azimuthal asymmetries and a global offset of the outer debris ring relative to the central star. Finally, we place stringent upper limits on the scattered light surface brightness of the outer ring. Using radiative transfer modeling, we find that it is impossible to account for all observed properties of the system under the assumption that both rings contain dust populations with the same properties. While the outer belt is in reasonable agreement with the expectations of steady-state collisional cascade models, albeit with a minimum grain size that is four times larger than the blow-out size, the inner belt appears to contain copious amounts of small dust grains, possibly below the blow-out size. This suggests that the inner belt cannot result from a simple transport of grains from the outer belt and rather supports a more violent phenomenon as its origin. We also find that the emission from the inner belt has not declined over three decades, a much longer timescale than its dynamical timescale, which indicates that the belt is efficiently replenished.
\end{abstract}



\section{Introduction}

Questions regarding the origin of the Solar System have pervaded the history of astronomy. Thorough studies of the architecture and dynamical history of our own Solar System \citep[e.g.,][]{horner14} and the discovery of thousands of extrasolar planets \citep{howard13} now provide an opportunity to test theoretical models and to piece together the story of planetary systems. Another natural laboratory of this history consists of debris disks, optically thin circumstellar disks composed mainly of dust surrounding Main Sequence stars \citep{zuckerman01}. In these systems, micron-sized dust grains, the only ones that can be probed directly, are continually blown out by radiation forces and replenished primarily via the grinding of planetesimals. Debris disks are observed at a rate of 15-20\% around Main Sequence stars, with increased frequency around younger stars \citep[][; Thureau et al., in prep.; Sibthorpe et al., in prep.]{su06, trilling08, carpenter09, eiroa13}. Therefore, they represent a crucial window into the planet formation process, each system providing a snapshot at a particular age and a chance to understand its dependency on other stellar parameters such as mass and metallicity. Understanding the diversity of debris disk systems is an important step towards placing our own Solar System in a global context and drawing a complete picture of planet formation.

Most debris disk systems are consistent with a steady-state collisional cascade within a gradually decaying reservoir of planetesimals, the canonical theoretical framework for these objects \citep{wyatt08}. However, this scenario is seriously challenged by systems showing very strong emission from hot/warm (i.e., close in) dust, especially if the system is relatively old \citep[$\gtrsim 100$\,Myr, i.e. past the epoch of terrestrial planet formation; for a recent review see][]{chambers11}. Instead, such systems are most likely transient remnants of a recent or ongoing event, either a rare catastrophic collision between very large planetesimals, a Late Heavy Bombardment-like phenomenon or a continuous replenishment via comets from a massive reservoir located much further out \citep{lisse12, kennedy13}. Dust belts located in the planet-forming region, i.e., within a few AU of the central star, can therefore arise from several paths and only detailed studies of these systems can inform on the likelihood of each scenario. Of particular interest is whether these particularly dusty warm debris disks are associated with a colder dust belt that could serve to feed dust grains to the innermost regions of the systems. Careful analysis of {\it Spitzer} mid-infrared spectroscopy has revealed that debris systems containing two separate dust belts, akin to the situation in our Solar System albeit with much more copious amounts of dust, are very common \citep[][Chen et al., submitted]{morales11}. The large physical separation between the two dust belts in these systems is suggestive of the presence of planetary bodies between them \citep{su13}, although it could also be explained by dust produced from planetesimals on initially high eccentricity orbits \citep{wyatt10}.

Debris disks are usually discovered by the infrared excess in the spectral energy distributions (SEDs) caused by the thermal emission of the dust. However, interpretations of the SED are fraught with an inherent degeneracy, in that a population of larger grains closer to the central star will exhibit the same total luminosity and equilibrium temperature as a population of smaller grains located further out. Indeed, spatially-resolved images of debris disks frequently reveal larger radii than would be inferred from pure blackbody emission \citep{rodriguez12, booth13, morales13}. Mid-infrared spectroscopy revealing silicate emission features can help break this degeneracy but is only available for a very limited number of targets \citep[e.g.,][]{chen06}. Thus, resolved imagery is essential to determine the dust properties, since it provides critical constraints on the location of the grains. Furthermore, spatially resolved images of debris disks can provide indirect evidence for the presence of planet-mass bodies that can interact gravitationally and shepherd the small dust grains \cite[e.g.,][]{kalas08, lagrange09}.

\ec\ is an F2, 1.4\,$M_\odot$, Main Sequence star located only 18.3\,pc away \citep{vanleeuwen07}, whose age has been estimated to be 1--2\,Gyr \citep{ibukiyama02, mallik03, vican12}. Significant infrared excess emission was first detected with {\it IRAS} \citep{stencel91}, and many subsequent studies over a broad wavelength range have targeted and analyzed this remarkable system. Most intriguingly, the SED of \ec\ exhibits a clear two-peaked excess, which is most easily explained by the fact that the circumstellar dust is distributed in two well separated belts. Indeed, \cite{wyatt05} presented sub-millimeter maps that resolved the system for the first time and showed that the cold dust belt, located $\sim$150\,AU from the central star, could not be responsible for the copious amounts of mid-infrared emission \citep[see also][]{beichman06, rhee07}. Instead, a second belt, located within $\lesssim$3\,AU of the host star \citep{smith08, smith09}, is the source of this emission at shorter wavelengths. Detailed analyses of the dust composition in this inner belt have been enabled by the spectroscopic capabilities of {\it Spitzer}, showing a rich mineralogy and the unambiguous presence of grains as small as 1\,\mic\ \citep{chen06, lisse12}. Most importantly, the mid-infrared excess produced by this dust belt is remarkably large considering the old age of the system, indicating that the gradual erosion via collisional grinding scenario is untenable \citep[e.g.,][]{wyatt07}. Throughout this paper, we interchangeably refer to the inner (outer) as the warm (cold) ring/belt.

As the 63$^{\rm rd}$ closest F-type star to the Sun \citep{phillips10}, \ec\ was included in the volume-limited Disc Emission via a Bias-free Reconnaissance in the Infrared/Sumillimetre survey (DEBRIS, Matthews et al., in prep.) which uses far-infrared continuum imaging with {\it Herschel} \citep{pilbratt10} to obtain more sensitive and higher resolution images of debris disk systems than previous observations in this domain. Early ``Science Demonstration" 100 and 160\,\mic\ images of \ec\ clearly resolved the outer dust belt \citep{matthews10} and warranted follow-up observations to take full advantage of the capabilities of {\it Herschel}. In this paper, we present new {\it Herschel} images of \ec\ at wavelengths ranging from 70 to 500\,\mic, as well as a new ground-based 850\,\mic\ map of the system obtained as part of the SCUBA-2 Observations of Nearby Stars survey \citep[SONS,][]{panic13}. The 70\,\mic\ image presented here is the highest resolution image of the system to date and it allows us to clearly disentangle the emission from both dust belts. We are thus able to determine precisely the geometry of the cold outer belt and to constrain the far-infrared emission properties of the warm inner belt. In turn, this enables us to constrain the dust properties in the outer belt and to compare them to those of the inner belt, providing crucial insight on the nature and current state of the system. We also present new visible and near-infrared {\it Hubble Space Telescope} ({\it HST}) images that allow us to place an upper limit on the amount of starlight that scatters off the outer belt.

The organization of this paper is as follows. In Section\,\ref{sec:obs} we present the new observations of \ec\ and the data reduction methods, as well as a re-analysis of its mid-infrared {\it Spitzer} spectrum. In Section\,\ref{sec:results}, we present the images as well as some basic analysis of both the SED, using modified blackbody models, and the images, using simple geometric models. In Section\,\ref{sec:model}, we show the results of a full radiative transfer model assuming a warm inner and cold outer belt, from which the dust properties in both dust belts are constrained, and we discuss the implications of our results in Section\,\ref{sec:discus}.


\section{Observations and data reduction}
\label{sec:obs}


\subsection{New {\it Herschel} observations}
\label{subsec:obs_herschel}

The data discussed in this work are distinct from those presented in \cite{matthews10}. The new observations were obtained with the Photodetecting Array Camera and Spectrometer \citep[PACS,][]{poglitsch10} in June and December 2011 (see Table\,\ref{tab:log}) using the default observing strategy of the DEBRIS survey (Matthews et al., in prep.), namely the ``mini scan map" mode at a scan rate of 20\arcsec/s and with scan legs of 3\arcmin. A single observation consists of two scans with a 40\degr\ difference in scanning angle, and this sequence was repeated once to obtain the desired depth. Because PACS observes in two channels simultaneously (70 or 100\,\mic\ in the ``blue" camera and 160\,\mic\ in the ``red'' camera), there are in total four separate scans at 70 and 100\,\mic, and eight at 160\,\mic. As \ec\ was strongly detected in the PACS Science Demonstration images, follow-up observations at longer wavelengths with the Spectral and Photometric Imaging Receiver (SPIRE) were triggered. The observations, obtained in January 2011, were conducted in the ``small map" mode. Simultaneous SPIRE maps were obtained at 250, 350 and 500\,\mic.

All raw observations were processed using the {\it Herschel} Interactive Processing Environment (HIPE, version 7.0). In particular, the data were high-pass-filtered to remove the marked low-frequency ($1/f$) noise present along the scanning direction in the PACS data using a spatial scale (1\arcmin) that is much larger than the extent of the target. This results in a net reduction of the flux of all sources as a consequence of the low-intensity wings of the PACS point spread function (PSF) that extend well beyond 1\arcmin. Thus all fluxes measured from the PACS images have been corrected appropriately. This low-pass filtering is unnecessary for SPIRE images. To take advantage of the inherently high level of redundancy of the PACS observations, we used the drizzle mode \citep{fruchter02} to generate maps at higher spatial resolution than the native pixel scale, leading to final maps with 1\arcsec/pix at 70 and 100\,\mic\ and 2\arcsec/pix at 160\,\mic, respectively. The SPIRE maps were constructed with the default pixel scale of 6, 10 and 14\arcsec/pix at 250, 350 and 500\,\mic, respectively.

After checking for mutual consistency in absolute pointing and gross source morphology, all independent PACS images were combined at both 100 and 160\,\mic\ to increase signal-to-noise. At 70\,\mic, where spatial resolution is highest and \ec\ is well-resolved, we carefully inspected all four images and noticed that one of the four maps was affected by a significant spike of correlated noise that coincided with the location of the target. We thus excluded this sub-map from the final combined map to improve the image quality at the expense of a modest loss in signal-to-noise.

\begin{figure}
\plotone{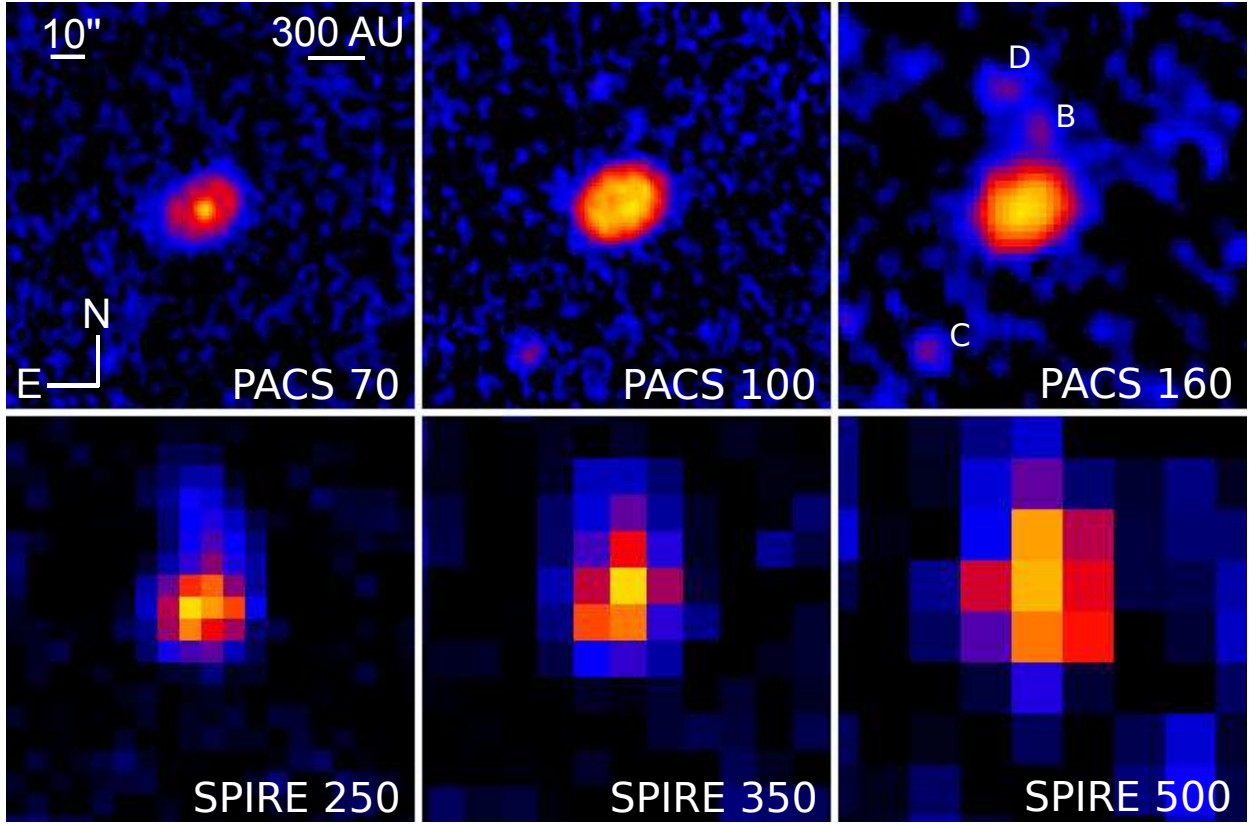}
\caption{Mosaic of all final PACS (top row) and SPIRE (bottom row) images of \ec. All images are shown on a square root stretch from zero to 125\% of the peak pixel brightness, which corresponds to 62, 45, 55, 23, 21 and 10 times the background noise in the 70, 100, 160, 250, 350 and 500\,\mic\ map, respectively. Three prominent background sources are labeled in the PACS 160\,\mic\ image. Sources B and C were first identified by \cite{wyatt05} and we add a new source D.\label{fig:images}}
\end{figure}

The full width at half-maximum (FWHM) of PACS images is about 5.6, 6.8 and 11.4\arcsec\ at 70, 100 and 160\,\mic\ \citep[note that there is some variability at 10\% level in the 70\,\mic\ FWHM; see, e.g.,][]{kennedy12b}, while that of SPIRE data is 18.2, 24.9 and 36.3\arcsec\ at 250, 350 and 500\,\mic, respectively. The sensitivity to point sources can be estimated by subtracting all detected sources in the map and placing a number of beam-sized apertures across the region of maximum data coverage. In the PACS images, these detection limits are 2--4\,mJy/beam (increasing towards longer wavelengths). In the SPIRE map, the background noise, estimated to be 6--8\,mJy/beam, is dominated by faint extragalactic sources. The final {\it Herschel} images are shown in Figure\,\ref{fig:images}.


\subsection{Ground-based sub-millimeter observations}
\label{subsec:obs_SCUBA-2}

\ec\ was observed as part of the SONS survey on the James Clerk Maxwell Telescope (JCMT). Observations were carried out using the SCUBA-2 camera \citep{holland13} at the primary waveband of 850\,\mic, where the telescope FWHM beam size is 13\arcsec. Individual observations were 30\,min in duration and a total of 14 were co-added together for a total integration time of 7\,hr. Most of the data were taken during the survey verification period for SONS in January 2012, although 1\,hr of integration was taken in May 2013. We adopted the constant speed ``DAISY" pattern observing mode, which is appropriate for compact sources and provides uniform exposure time coverage in the central 3\arcmin-diameter region of the field. The atmospheric opacity was monitored in real-time using a line-of-sight water vapor monitor, and conditions were generally good, with typical zenith sky opacities of 0.4 at 850\,\mic. However, the sky conditions were too poor to yield usable results at the shorter SCUBA-2 waveband of 450\,\mic. 

The data were reduced using the Dynamic Iterative Map-Maker within the STARLINK SMURF package \citep{chapin13}. The technique of ``zero-masking" of the astronomical signal was adopted. Using this method the map is set to zero beyond a radius of 90\arcsec\ (for this case) of the field center for the majority of the iterative part of the map-making process. Such a constraint helps to suppress the large-scale ripples that can produce ringing and other artefacts in the final image. The result is flatter looking maps over the case in which the data are high-pass filtered \citep[e.g.,][]{panic13} and where the mean measured signal in the vicinity of the source (but not including the source) is close to zero. The use of such masking has resulted in improved signal-to-noise of detected structure as well as more accurate flux measurements from aperture photometry compared to previous data reduction methods. 

\begin{figure}
\plotone{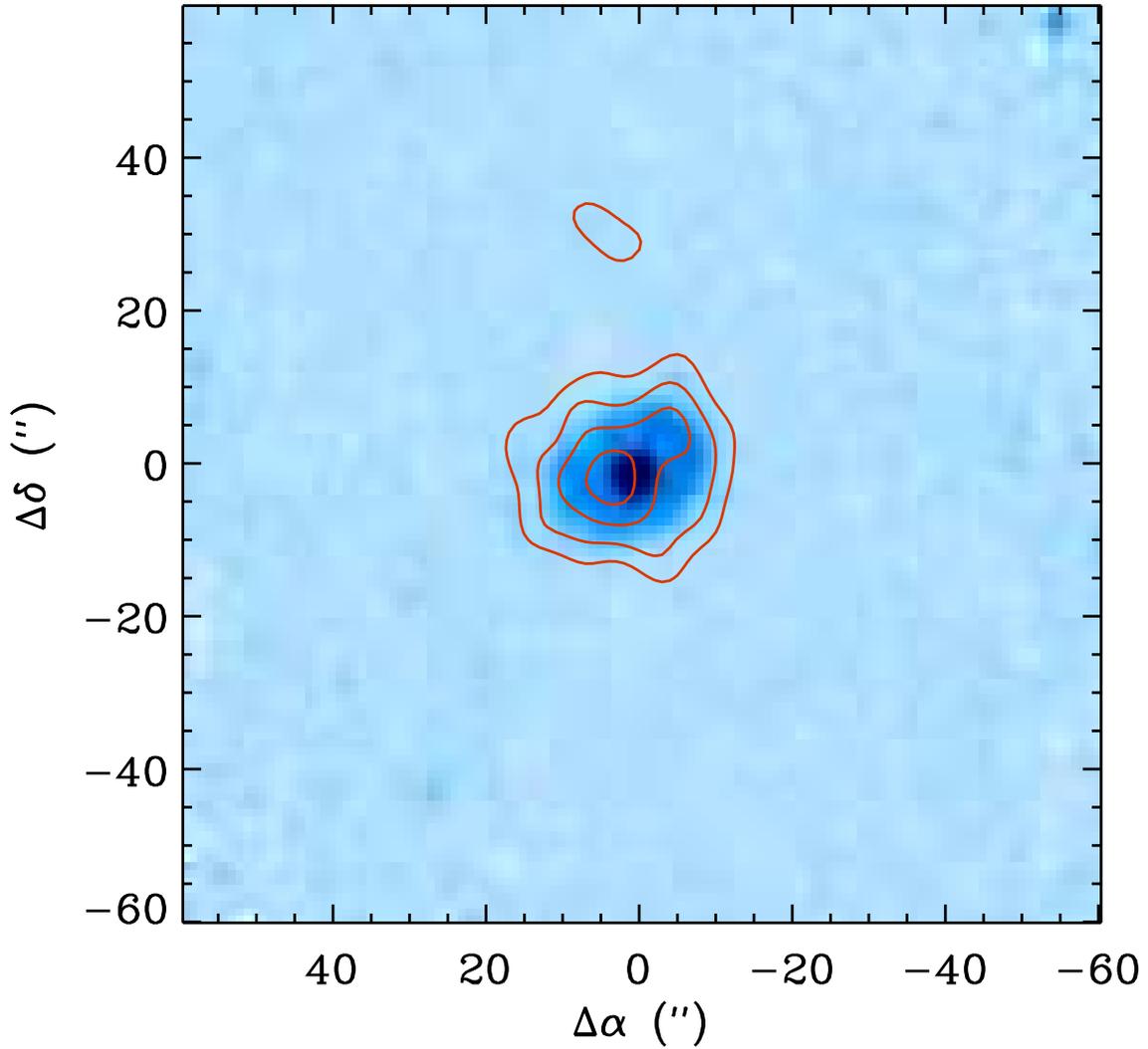}
\caption{SCUBA-2 850\,\mic\ map (4, 6, 8 and 10$\sigma$ contours) overlaid on the 70\,\mic\ PACS image (colorscale). The two images were aligned using the absolute coordinates provided by each instrument. The absolute pointing uncertainties of both {\it Herschel} and SCUBA-2 are about 2\arcsec\ each.\label{fig:SCUBA-2}}
\end{figure}

The data were flux calibrated against the primary calibrator Uranus and also secondary sources CRL\,618 and CRL\,2688 from the JCMT catalog \citep{dempsey13}. The estimated calibration uncertainty was 5\%, based simply on the spread of derived flux conversion factors from the calibrator observations. The level of background noise, 1\,mJy/beam, was estimated from the standard deviation of multiple measurements of the flux within identical 40\arcsec\ apertures, placed 10\arcsec\ apart, over the central 3\arcmin\ circular region (but avoiding the source). The final image, shown as contours in Figure\,\ref{fig:SCUBA-2}, has been smoothed with a 7\arcsec\ FWHM Gaussian to improve the signal-to-noise. 


\subsection{Visible and near-infrared imaging}
\label{subsec:obs_hst}

Multi-wavelength images of \ec\ were obtained using both the Advanced Camera for Surveys (ACS) and the Near-Infrared Camera and Multi-Object Spectrometer (NICMOS) cameras on board {\it HST} via program 10244 (PI: M. Wyatt). The ACS observations utilized the coronagraphic mode of the High Resolution Channel (HRC), along with its 1\farcs8 diameter occulting spot, yielding a 29\farcs2$\times$26\farcs2 field of view at a pixel scale of 0\farcs028 x 0\farcs025. Target exposures in the F435W, F606W, and F814W filters were obtained at two different {\it HST} roll angles, separated by 28\degr, to better facilitate the separation of PSF artifacts from bona-fide disk scattered light features. The near infrared coronagraphic observations of \ec\ were obtained with the NICMOS-2 camera, a 256$\times$256 HgCdTe array with a pixel scale of 0\farcs076, offering a full field of view of 19\farcs5. The target was imaged at the predefined low-scatter point of the coronagraphic system, behind the 0\farcs6 diameter hole. Coronagraphic observations were obtained in the F110W, F160W, and F171M filters at a single roll angle. Repeated sequences of images were obtained in MULTIACCUM mode to achieve the total exposure for each of our observations. In addition to observations of \ec, we also obtained data on HD\,105452 (with ACS, using a single roll) and HD\,132052 (with NICMOS) with the same observing strategy, to serve as PSF references. A complete log of observations is presented in Table\,\ref{tab:hst}. All datasets were reduced using standard routines from the {\it HST} pipeline (STSDAS), using the latest available calibration files to correct for dark current, bias, flat field and electronic noise effects and to reject cosmic rays.

PSF subtraction of our ACS data followed techniques commonly used in analysis of coronagraphic data \citep[e.g.][]{clampin03, wisniewski08}. We normalized and aligned the non-distortion corrected images of our PSF reference star to \ec\ in an iterative manner using a cubic convolution interpolation function. Residual alignment errors in the resultant best subtraction are on the order of 0.05 pixels \citep{gonzaga13} and our normalization is accurate to $\pm$2\% \citep{clampin03}. Following the subtraction of a fully registered and scaled PSF, our \ec\ data were corrected for geometric distortion and rotated into a common orientation via the {\it MultiDrizzle} routine. We used {\it Synphot}, the synthetic photometry package within STSDAS, to determine the correction factor needed to calibrate our data to an absolute photometric scale. All PSF-subtracted, distortion-corrected images were normalized to the synthetic flux of \ec\ and co-added to increase the signal to noise ratio of our data. Building on our observing strategy, we also explored using one roll angle image of \ec\ as a PSF template for the other roll angle image. This PSF subtraction technique has been previously explored for use in other ACS/HRC coronagraphic imaging programs, including successful implementation in situations where the disk is at moderate inclination \citep{krist05, krist10}. Following optimal image registration and distortion correction, we rotated the images into a common orientation, and combined them to increase the effective signal to noise ratio. Unbeknownst to us at the time these data were obtained, the object selected as primary PSF reference star (HD\,105452) has an apparent companion\footnote{Analysis of archival near-infrared images of the system from CFHT and ESO (taken in 2003 and 2008, respectively) indicates that this object is actually a background star.} at a projected separation of 2\farcs22 and position angle 255\fdg4 which significantly degrades our ability to obtain clean residuals in the F814W filter data. This problem is much less acute in the F435W and F606W filters. Thus, we also retrieved archival data on three other targets with similar spectral type and apparent broadband colors and observed with the same set-up. We performed the same reduction and PSF subtraction steps as described above to obtain independent PSF subtracted images of \ec.

A similar iterative, cubic convolution interpolation function was used to scale and register the NICMOS images of \ec\ and the PSF reference star. We estimate that the accuracy of the alignment is $<$0\farcs1. Despite of our best efforts, the use of HD\,132052 as a PSF template routinely resulted in significant residuals appearing in the PSF-subtracted images of \ec. \cite{schneider01} provide a detailed discussion of the various {\it HST} thermal instabilities which likely cause such mismatches. To better search for evidence of spatially resolved circumstellar material around \ec, we identified several F-type stars in the {\it HST} archive that have been obtained with the same instrument configuration as for our \ec\ observations. These stars can also serve as PSF template stars and thus we processed these archival data in a similar manner as described above, including the scaling and registering process with \ec. Finally, all of the PSF-subtracted images of \ec\ were corrected for the geometric distortion which is inherent to all NICMOS data. 


\subsection{Mid-infrared spectroscopy}
\label{subsec:obs_irs}

A mid-infrared spectrum of \ec\ was obtained using the Infrared Spectrograph \citep[IRS,][]{houck04} instrument aboard {\it Spitzer}. A first analysis of this dataset was published by \cite{chen06}, which was superseded by the analysis of \cite{lisse12} who used improved extraction methods. Here, we make use of the spectrum from the latter study, with one modification regarding the absolute flux calibration of the spectrum. Instead of only using mid-infrared fluxes from the IRS pick-up camera at 16 and 22\,\mic\ and the MIPS 24\,\mic\ photometry to establish the absolute photometry of the IRS spectrum, we include in the calibration the {\it AKARI} 9 and 18\,\mic, {\it WISE} 12 and 22\,\mic\ and MIPS 24\,\mic\ broadband fluxes (listed in Table\,\ref{tab:sed}). The relatively wide bandpasses of these different filters enable us to calibrate the spectrum across most of the IRS wavelength range, thereby providing a more robust estimate for the absolute flux calibration of the spectrum.

We find that the spectrum processed by \cite{lisse12} needs to be scaled by a factor of 0.86$\pm$0.04 to match with the broadband fluxes, with no significant trend as a function of wavelength. This scaling factor is consistent with the $\approx10\%$ precision on the absolute flux calibration quoted by the authors of that study. We apply this correction factor in our analysis, noting that it results in somewhat weaker excess in the IRS range than previously found after subtraction of our best estimate of the stellar photosphere (see Section\,\ref{subsec:modbb}). At wavelengths longer than about 8\,\mic, the excess flux is decreased by a factor of $\sim$2 but its shape is essentially unaffected. The excess at shorter wavelengths is reduced more substantially but we note that the several percent uncertainties estimated for the absolute scaling of the IRS spectrum and for the stellar photospheric emission (Section\,\ref{subsec:modbb}) prevent us from probing this excess with adequate precision. A more detailed analysis of these factors is required to establish the level of excess at wavelengths $\lesssim8\,$\mic, which is beyond the scope of the present work.


\section{Observational results}
\label{sec:results}


\subsection{\ec\ at far-infrared wavelengths}
\label{subsec:basic_res}

\ec\ is strongly detected in all three PACS images, with integrated signal-to-noise ratios of 55, 73 and 58 at 70, 100 and 160\,\mic, respectively. Total fluxes were estimated using aperture photometry (with a 20\arcsec\ aperture radius) and applying appropriate aperture corrections, leading to the fluxes listed in Table\,\ref{tab:sed}. At all three wavelengths, \ec\ is unambiguously spatially resolved. At 70\,\mic, the morphology of the system consists of a central core surrounded by an elliptical area of fainter emission extending out to a maximum of about 15\arcsec\ from the central source along the SE--NW direction. The 100 and 160\,\mic\ images show an essentially flat-topped elliptical structure whose size and orientation matches the maximum extent of the 70\,\mic\ emission; no central core is detected in either image. 

\begin{figure}
\plotone{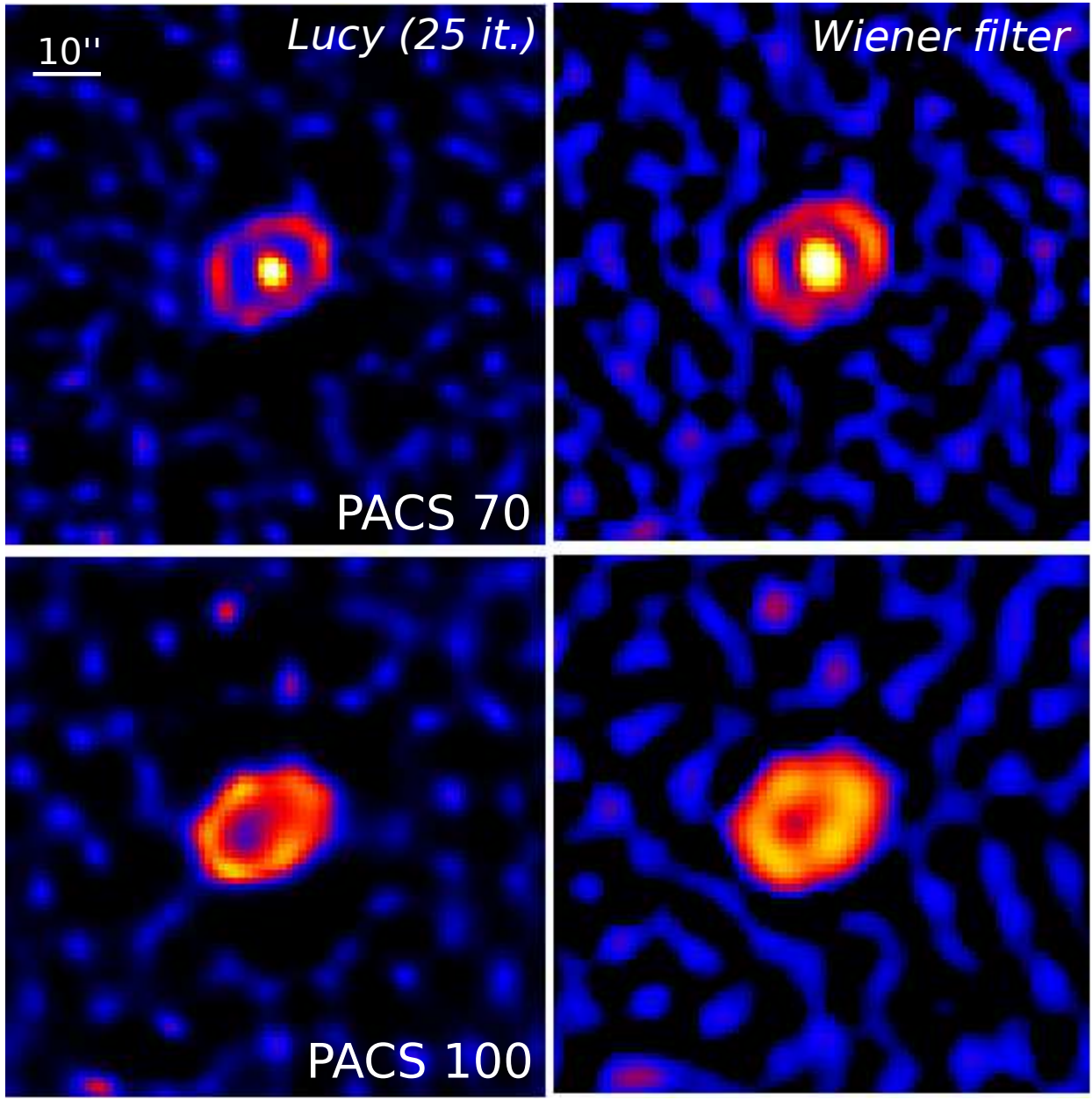}
\caption{Deconvolved 70 (top row) and 100\,\mic\ (bottom row) images of \ec\ using the Lucy algorithm (left column) and Wiener filtering (right column). The field of view of the maps is 80\arcsec\ and the orientation is the same as in Figure\,\ref{fig:images}. Note that sources B, C and D at 100\,\mic\ appear more prominently in the deconvolved images, but that their significance is not enhanced by the deconvolution process.\label{fig:deconv}}
\end{figure}

The simplest interpretation of the PACS images is that the system consists of a central source surrounded by an inclined ring that is also responsible for the sub-millimeter emission, with a contribution of the central source to the total flux declining steeply between 70 and 100\,\mic\ to become almost negligible. After taking into account the background emission introduced by the outer ring, the central source has a FWHM that is consistent with the PACS beam at 70\,\mic\ and thus is unresolved, as further demonstrated in Section\,\ref{subsec:geom}. To confirm this geometrical configuration, we have deconvolved the 70 and 100\,\mic\ PACS images using the IRAF implementations of the Richardson-Lucy algorithm (for 25 iterations) and Weiner filtering (see Figure\,\ref{fig:deconv}).  Although we caution against over-interpretation of deconvolved images, these can be useful in confirming findings that are first identified in the corresponding raw images. The deconvolved images clearly reveal the outer ring at both wavelengths and its clean separation from the central point source at 70\,\mic. They further suggest a brightening of the outer ring ansae at 70\,\mic\ as expected from the geometry-induced increase in optical depth, although the significance of this result is marginal at best. Overall, the morphology of \ec\ in the PACS images is strongly reminiscent of that of HD\,207129 \citep{marshall11}, with the most immediate difference being that the outer ring in \ec\ is seen at an inclination that is further away from edge-on.

\begin{figure}
\plotone{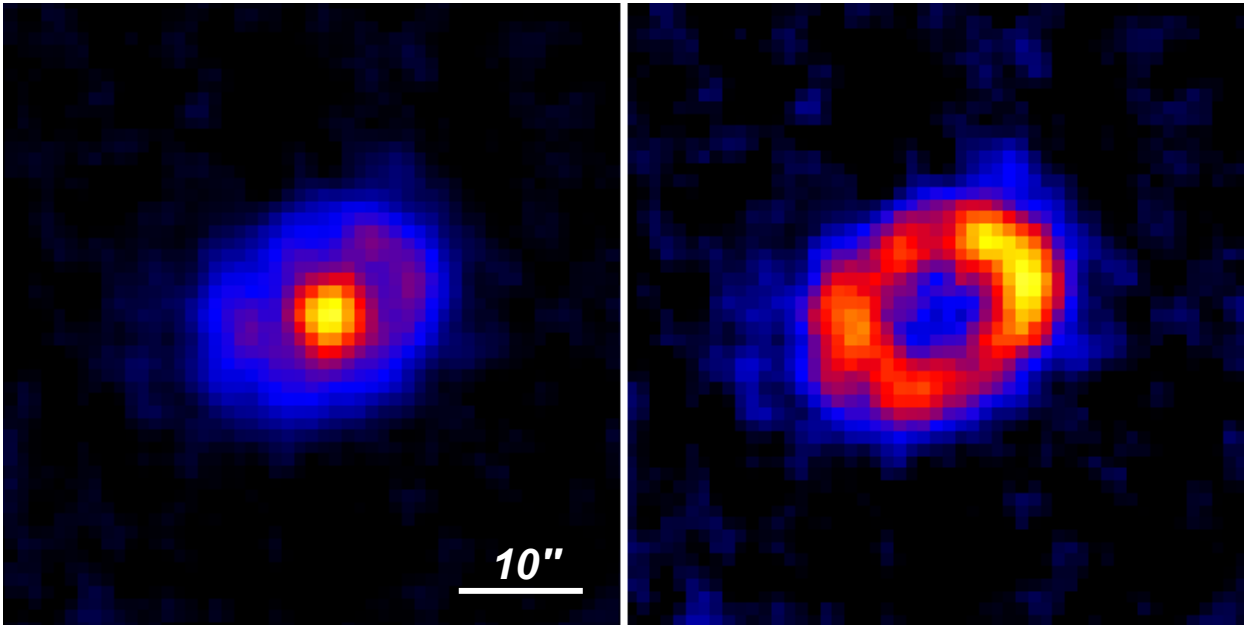}
\caption{Close up on the 70\,\mic\ image of \ec\ before (left) and after (right) subtraction of a PSF scaled to the best-fit estimate of the combined flux of the star and inner ring component. Both images are shown on a linear stretch from zero to 125\% of the peak pixel brightness. The signal-to-noise ratio per pixel varies from 10 to 20 along the ring in the PSF-subtracted map.\label{fig:psfsub}}
\end{figure}

Subtracting a point source with the expected photospheric flux ($F_\nu^\star\approx35$\,mJy) from the 70\,\mic\ image  leaves a bright central source, indicating that there is unresolved excess emission at the star position, presumably associated with the inner ring of warm dust. To estimate the total flux of the central source (i.e., the sum of the stellar and inner ring flux), we aligned and scaled a reference PSF (Figure\,\ref{fig:psfsub}). Out of 11 available PSFs, we selected an image of $\beta$\,And (OID\,1342212508) which presented the best match to the core FWHM measured in \ec\ image. We note that the low intensity ($\approx$10\% of the peak) trefoil artefact of the PACS PSF may introduce a small additional uncertainty in the subtraction. The inclination and compact size of the ring also introduces uncertainty in the PSF subtraction process. Using two extreme cases of a zero-flux central pixel on one hand and an maximally smooth residual map inside of the ring on the other hand (which yield fluxes of 79 and 61\,mJy for the unresolved source, respectively), we estimate that the central point source has a 70\,\mic\ flux density of $70\pm5$\,mJy. Thus the emission from the warm dust at 70\,\mic\ is comparable to that of the star itself. Note that the absolute flux calibration uncertainty of PACS (2.6\%\footnote{http://herschel.esac.esa.int/twiki/pub/Public/PacsCalibrationWeb/pacs\_bolo\_fluxcal\_report\_v1.pdf}) is significantly smaller and thus the dominant source of error is the PSF subtraction itself. This estimate is confirmed by the analysis of the deconvolved images, which show that the central point source accounts for about 30\% of the total flux. At 100\,\mic, no central point source is detected, even in the deconvolved image. Subtracting a scaled PSF from the image until the central pixel has zero flux leads to a conservative upper limit of 45\,mJy for the central source. For comparison, the predicted photospheric flux at 100\,\mic\ is 17\,mJy.

\begin{figure}
\plotone{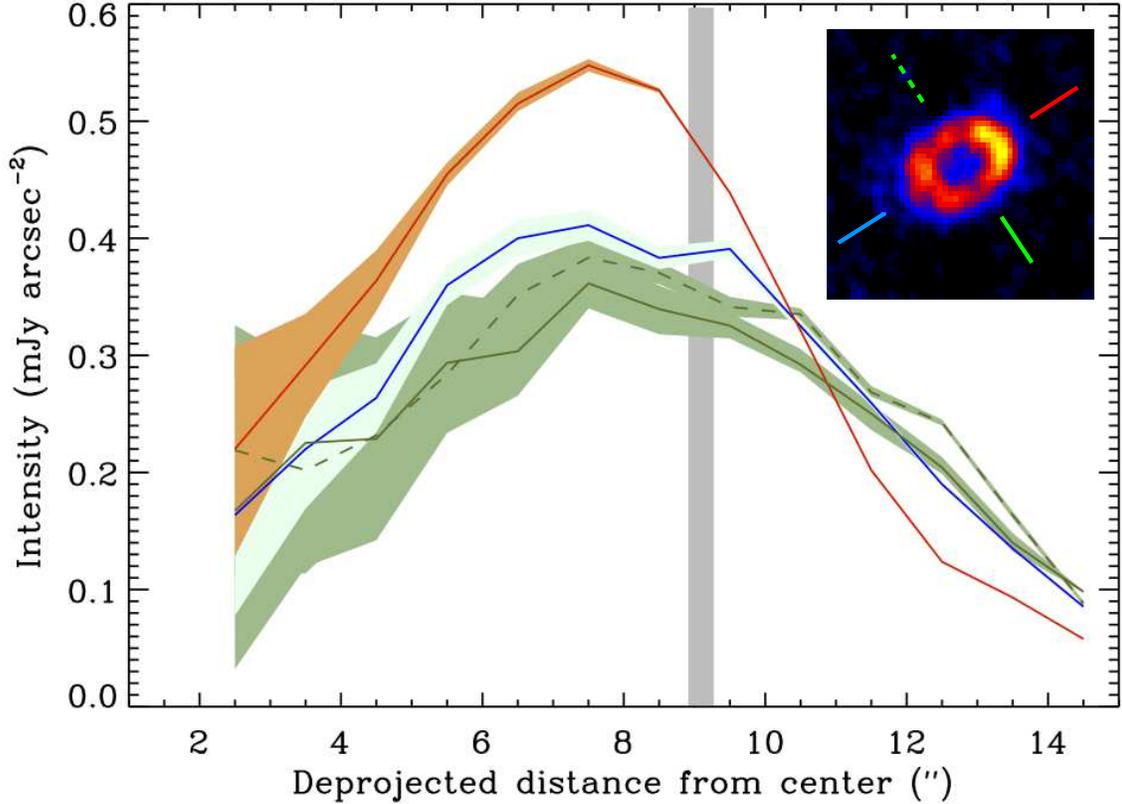}
\caption{Surface brightness radial profile for the outer ring of \ec\ as estimated from the core-subtracted 70\,\mic\ image. The profiles are computed as the median profiles within four separate wedges, each with a half opening angle of 30\degr\ and centered along either side of the major and minor axes. The inset image indicates the central PA used to define each wedge. The shaded areas surrounding each curve represent the range of ring profiles obtained from using the ``minimum" and ``maximum" possible PSF subtraction, highlighting the increasing uncertainty introduced by PSF subtraction at short distances from the star. The horizontal axis represents the de-projected distance to the central source assuming an inclination of 47\degr\ and a major axis PA of 116\degr\ (Section\,\ref{subsec:geom}). The gray bar indicates the position of the ring mean radius as derived from our radiative transfer modeling (Section\,\ref{subsec:model_outer}); it is further out than the peak of the curves as a consequence of azimuthal smearing and PSF convolution. \label{fig:ringprofile}}
\end{figure}

The 70\,\mic\ image of the outer ring (after subtraction of the central core) shown in Figure\,\ref{fig:psfsub} provides valuable information on its azimuthal and radial structure. The NW ansa has a higher peak surface brightness and is somewhat sharper than the SE one, as illustrated in Figure\,\ref{fig:ringprofile}. The radial profiles of the ring appear marginally resolved in all directions, even towards the sharper NW ansa. This does not seem to be confirmed by our deconvolution attempts, however. Our modeling (see below) supports the conclusion that the ring is not resolved radially in these observations. Higher resolution observations, which would be particularly helpful in reducing confusion with the central point source, are needed to study directly the radial and azimuthal structure of the outer ring. We note, however, that the SE and NW halves of the outer ring (split along the minor axis) have integrated 70\,\mic\ fluxes whose ratio is $0.96^{+0.02}_{-0.03}$. In other words, there is no significant left/right asymmetry in the integrated brightness of the outer ring at that wavelength.

At 100 and 160\,\mic\, several faint sources are detected in the vicinity of \ec\ (see Figures\,\ref{fig:images} and \ref{fig:deconv}). Two of them, sources B and C, have been detected at 450 and 850\,\mic\ by \cite{wyatt05} and we follow their nomenclature, adding a source D located further North of \ec\ than source B. We estimated the 100 and 160\,\mic\ fluxes of all three background sources using small apertures and applying appropriate aperture corrections; we also estimated their location relative to \ec\ by measuring the centroid of each source in these apertures. The relative astrometry and fluxes of the background sources are summarized in Table\,\ref{tab:galaxies}. Given its high proper motion, \ec\ has moved $\approx4$\arcsec\ relative to fixed background sources during the almost 9\,yr time difference between our PACS observations and the earlier SCUBA observations of \cite{wyatt05}. While such a displacement could be easily measurable considering the signal-to-noise ratio of both observations, the wavelength-dependent morphology of both the background sources and \ec\ introduce significant uncertainties in both datasets. Taking astrometric uncertainties at face value (i.e., only including centroiding precision and instrumental astrometric calibration), source B is consistent with a fixed background object whereas source C is more consistent with a co-moving object. However, realistic estimates of the total uncertainties are large enough that the data at hand do not allow to be conclusive yet. Observations in the next few years should shed definitive light on the nature of these sources.

In the SPIRE images, \ec\ is confused with sources B and/or D, whose contribution to the total flux increases toward longer wavelengths. To reduce uncertainties associated with confusion, total fluxes from the SPIRE maps were derived using PSF fitting. The contribution of sources B and D is marginal at 250\,\mic\ (on the order of 10--20\%), but becomes increasingly strong at longer wavelengths. We use fluxes from PSF fitting as upper limits to the flux of \ec\ at 350 and 500\,\mic.

The SCUBA-2 850\,\mic\ map shows a clear detection of \ec, which is furthermore spatially resolved and marginally asymmetrical. In short, the source appears to consist of two separate peaks (detected at the 10.5 and 8$\sigma$ level, respectively, the brightest of which is to the SE) aligned along the same position angle (PA) as the major axis of the ring seen in the PACS images. The separation between the peaks is about 9\arcsec, i.e., slightly smaller than the image resolution. This is broadly consistent with the previous SCUBA map obtained by \cite{wyatt05}, although our new observations are of somewhat higher resolution and signal-to-noise. The SE-NW asymmetry of the ring, which was not significant in the original SCUBA map, is marginally significant and still requires confirmation from higher signal-to-noise data, for instance with ALMA. The total flux from \ec\ in the SCUBA-2 map (see Table\,\ref{tab:sed}) was estimated using a 40\arcsec-radius aperture centered on the peak of emission. The dominant source of uncertainty is the absolute calibration of SCUBA-2 rather than the statistical noise in the map. Besides \ec, source D is marginally detected in our 850\,\mic\ SCUBA-2 image, with a flux of 4$\pm$1\,mJy. Sources B and C are not detected, suggesting that their flux density is weaker than estimated by \cite{wyatt05}. All three sources have SEDs that peak at a wavelength of 200--300\,\mic, typical of high-redshift, dust-rich galaxies. We thus believe that the sources detected in the {\it Herschel} and SCUBA maps are most likely extragalactic objects.

\begin{figure}
\plotone{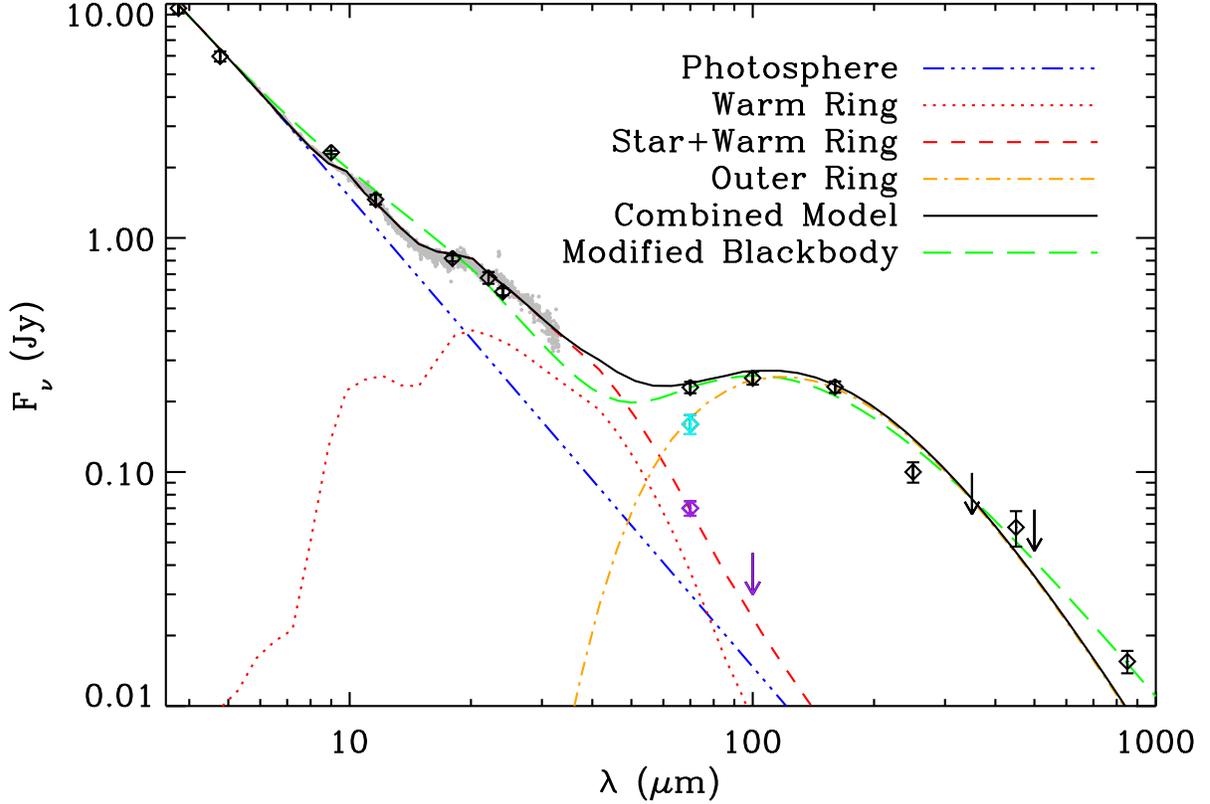}
\caption{Observed SED of \ec\ compared with the best fitting model from our full radiative transfer calculations (Section\,\ref{sec:model}) and modified blackbody model (Section\,\ref{subsec:modbb}). The various components of the full radiative transfer best fit model are shown. The black diamonds and gray circles represent the broadband photometry and IRS spectrum of the system, respectively. At 70 and 100\,\mic, we show the contribution from the outer belt alone and the central point source (star and inner belt) with cyan and purple symbols, respectively.  \label{fig:sed_mcfost}}
\end{figure}

The spectral index, defined as $F_\nu \propto \nu^\alpha$, is very shallow at the longest wavelengths compared with other debris disks \citep{roccatagliata09, panic13}. Combining the 450\,\mic\ SCUBA flux with the new 850\,\mic\ SCUBA-2 flux, we derive $\alpha = 2.1 \pm 0.3$, marginally flatter than the value derived by \cite{wyatt05} and consistent with a Rayleigh-Jeans tail. Including shorter wavelength fluxes in the fit (from SPIRE and PACS) further flattens the spectral index. The complete SED is shown in Figure\,\ref{fig:sed_mcfost}.


\subsection{Modified blackbody fits to the SED of \ec}
\label{subsec:modbb}

As a first step in modeling the SED of \ec, we estimated the stellar properties by using the available photometry of the system for $\lambda < 5$\,\mic\ (see Table\,\ref{tab:sed}). We adopted the NextGen atmospheric models \citep{allard97} and determined that the best fitting model has $T_{eff} = 7000$\,K and $R_\star = 1.5 R_\odot$ (for a total luminosity of 4.9\,$L_\odot$). We estimate that the precision on the stellar luminosity with this method, which stems from photometric uncertainties and slight mismatches between filters, is on the order of 4\%, which is sufficient to study the thermal emission of the debris rings. No foreground extinction is needed to obtain a good fit ($A_V \leq 0.05$\,mag). This is in good agreement with previous estimates of the stellar properties. 

In order to get a sense of the basic properties of the SED, we construct a simple model of the stellar flux plus the emission of two independent dust components, representing the warm and cold rings, respectively. Each component was modeled assuming modified blackbody dust emission at a single temperature (narrow ring) and and a wavelength-dependent absorption efficiency defined by
\begin{equation}
Q_\lambda = 1 - exp\left[-\left(\frac{\lambda_0}{\lambda}\right)^{\beta} \right]
\end{equation} 
\citep[e.g.,][]{williams04}. There are four parameters for each dust belt: temperature, total disk luminosity, grain efficiency $\beta$, and $\lambda_0$ for a total of eight parameters. We limited the $\beta$ parameter to the physically plausible [0,2] range, but left all other parameters unconstrained. The best models and uncertainties were found using a Monte Carlo Markov Chain (MCMC) algorithm \citep{goodman10}. If we assume that the emission is from grains of a single (characteristic) size and density 2.9\,g/cm$^3$ (see Section\,\ref{sec:model}), the model parameters can be reinterpreted in terms of the grain size ($a = \lambda_0 / 2\pi$; throughout this paper "grain size" refers to the grain radius) and total dust mass.

The parameters for the warm component were first fit to the five broadband fluxes between 8\,\mic\ and 40\,\mic\ in addition to the flux of the warm component at 70\,\mic. The models were also forced to remain below the 3$\sigma$ upper limit at 100\,\mic, although this constraint only serves to prevent the chain from converging to physically implausible models. In a second step, the parameters for the warm component were fixed to their best estimates to model the cold component to the remaining points longward of 70\,\mic. This sequential approach to fitting the two components, which avoids probing a parameter space of high dimensionality, is supported by the nearly orthogonal datasets probing each dust ring: the two components are cleanly separated in the 70\,\mic\ image and the warm belt contributes little flux at 100\,\mic\ and beyond. The resulting fit is shown in Figure~\ref{fig:sed_mcfost}. 

The warm ring models converged to a grain temperature of $418\pm16$\,K and a total fractional luminosity of $L_{disk}/L_\star = (3.3\pm0.2) \cdot 10^{-4}$ in agreement with previous fits \citep{wyatt05, matthews10, lisse12}, albeit with much tighter uncertainties on both parameters. The other fit parameters are not as tightly constrained (Table\,\ref{tab:modelpars}), in large part because of correlations between parameters. For instance, the lower limit on the total dust mass is $5 \cdot 10^{-7}\,M_\oplus$. Notably, the derived $\lambda_0$ is suspiciously close to $\lambda_{max}$ (the longest wavelength used in the fit) and may thus be considered a lower limit. Because of this, it is surprising at face value that we can constrain somewhat the grain efficiency parameter $\beta$. Furthermore, even though the apparent slope cast by the 20\,\mic\ and 70\,\mic\ broadband fluxes and the 100\,\mic\ upper limit is nearly parallel to the star's SED, our modeling readily excludes a solution with $\beta = 0$, which would be characteristic of perfect blackbody emission. The relatively tight error ranges on the short wavelength fluxes are such that they anchor the wavelength of the peak emission, hence the dust temperature. Since the Planck function reaches its power law asymptotic behavior at wavelengths $\lambda \gtrsim 3 \lambda_{peak}$, a pure blackbody function cannot simultaneously match the peak wavelength and the 20--70\,\mic\ slope, thus requiring the opacity law to compensate and leading to the model parameters listed above. Regardless, the nature of this simple model leads to an unavoidable failure to encompass the mid-IR spectral features in the IRS spectrum, casting doubt upon the exact physical relevance of its parameter values. A more detailed analysis is presented in Section\,\ref{sec:model} where a more physical model of the dust emission is used instead.

Regarding the cold ring, the constrained parameters were in agreement with previous fits with a temperature of $42\pm1$\,K and a fractional luminosity of $(2.2\pm0.1) \cdot 10^{-5}$. The improved precision for these parameters stems from the much denser wavelength coverage, particularly around the wavelength of peak emission. The other two parameters were constrained only to the extent that they resulted in a model with an opacity behavior of a perfect blackbody (i.e., a constant $Q_\lambda$), in agreement with the spectral index found in Section\,\ref{subsec:basic_res}. One possibility is that $\beta \approx 0$, in which case the model is explicitly a blackbody and $\lambda_0$ is unconstrained by definition. Alternatively, it could be that $\lambda_0 \gg \lambda_{max}$ (specifically, we find that $\lambda_0 \geq 3$\,mm), and the dust emission again obeys a blackbody to first order. In this case, both $\beta$ and the dust mass are unconstrained since the latter is tightly correlated with the grain size. Using $\lambda_0 = 3$\,mm, we derive a lower limit to the dust mass of $8 \cdot 10^{-3} \,M_\oplus$. Irrespective of the modeling details, the outer belt must contain a much larger surface area (and thus mass) of dust than the inner belt since its fractional luminosity is only an order of magnitude lower despite being ten times colder \citep[see also][]{lisse12}.


\subsection{Geometrical models}
\label{subsec:geom}

Based on the qualitative analysis of the PACS 70\,\mic\ image, we carried out a variety of geometric tests using various permutations of a model consisting of a central source encircled by an inclined geometrically flat ring. The central source combines the emission from the star and the warm dust ring, which are assumed to be concentric. Model images of the outer ring were generated at a resolution ten times the natural pixel resolution, re-binned, and convolved with the PSF. A total $\chi^2$ was computed by summing the uncertainty-weighted difference between model and observations over all pixels in a 30\arcsec$\times$30\arcsec\ window centered on \ec. The uncertainty per pixel is assumed to be uniform throughout this small sub-map, given the almost uniform coverage resulting from our observing strategy. The pixel uncertainty is estimated via the standard deviation of neighboring background pixels, multiplied by a factor of 3.6 to account for correlated noise introduced by the drizzle method and our choice of a final pixel scale that is smaller than the native pixel scale \citep{fruchter02}. This multiplicative factor has been shown to produce reasonable parameter uncertainties when compared to other empirical approaches \citep[e.g.][]{kennedy12}. For each geometrical model, the parameter space was explored using an MCMC algorithm. 

The base set of parameters for the models were the ratio of integrated fluxes between the central source and the outer ring, the mean radius ($R_{mid}$), width ($\Delta R$), PA of the major axis (measured East of North), and inclination ($i$) of the outer ring, as well as a two-dimensional positional offset between the outer disk and the central point source. Since the outer ring appears radially unresolved, we first adopt a default ring width of 10\,AU. We use this as our reference model, although we relaxed these assumptions in separate tests discussed below. We note that the MCMC-derived parameter uncertainties do not take into account intrinsic shortcomings of the models, such as the possibility of azimuthal asymmetries in the outer ring. However, we find that the resulting parameter uncertainties are comparable to, or larger than, the dispersion of values obtained for the various models explored, suggesting that they are realistic.

From our base model, we conclude that the mean radius of the outer ring is 164$\pm$2.5\,AU (9\farcs0$\pm$0\farcs1), and its inclination and PA are 47$\pm$1\degr and 117$\pm$2\degr, respectively. The precision on these parameters is greatly improved compared to the estimates of \cite{wyatt05} (150$\pm$20\,AU, 45$\pm$25\degr and 130$\pm$10\degr, respectively) owing to the excellent spatial resolution of {\it Herschel} at 70\,\mic. The semi-major axis we estimate here is larger than would be estimated directly from the location of the peak surface brightness as a consequence of the convolution of the inclined ring with the relatively broad PACS PSF, which smears its maximum radial extent (see Figure\,\ref{fig:ringprofile}). Furthermore, we find that the inner source (star + inner ring) contributes 30$\pm$1\% of the total flux at 70\,\mic. The predicted blackbody temperature at the outer ring distance from the central star is 32\,K, so that the observed dust temperature is about 1.45 times higher than for perfect blackbody grains, suggesting that some dust grains are smaller than 10\,\mic\ or so. Equivalently, the ring radius derived from the PACS image is a factor of $\Gamma = 2.1$ larger than would be inferred from the dust temperature, in line with results for debris disks around solar-type and intermediate-mass stars \citep[e.g.,][]{morales13}. We have adapted the model developed in \cite{booth13}, which assumes that the dust distribution extends down to the blow-out size but is devoid of smaller grains, to the observed stellar properties and ring radius of \ec, and derived a predicted value of $\Gamma_{model} = 3.2$. The observed lower value of $\Gamma$ suggests a deficit of grains around the blow-out size. We discuss this in more detail after we develop a more complete radiative transfer modeling of the dust belt.

Allowing the ring width to vary always led this parameter to the smallest allowed value with a $3\sigma$ upper limit of 60\,AU. The radial extent of the ring is explored in more detail in Section\,\ref{sec:model}. We also allowed the central source to be spatially resolved, assuming that it has an intrinsic Gaussian surface brightness profile. The best fit intrinsic Gaussian FWHM is 0\farcs9$\pm$0\farcs2. While this appears as evidence for a marginally resolved source, we note that this finite size only increases the apparent FWHM of the image by 0\farcs05 once convolved by the PSF. Since this  is less than the uncertainty on the intrinsic width of the PSF, we do not believe that it is significant and treat the core component as spatially unresolved. From the quadratic difference between the FWHM of the broadest available PSF and that of the \ec\ core component, we place an upper limit of 45\,AU on the intrinsic diameter of the latter. Finally, we explored whether the PACS image supports the presence of a third component to the model, with a dust component extending from the central source to the outer ring with a flat surface brightness profile. We do not significantly detect this component and place a 3$\sigma$ upper limit on its surface brightness of 1\% of the peak surface brightness from the outer ring. This results in an upper limit on the integrated flux of this intermediate component of 42\,mJy at 70\,\mic.

When an offset between the central source and the center of the outer ring is allowed, a non-zero offset is systematically preferred, with the ring center being located 0\farcs9$\pm$0\farcs2 (about 16\,AU) North of the location of the central source, independently of the specific geometrical model adopted. This $\approx$10\% offset appears to be significant, yet could be influenced by departures from the simple models we considered. Local brightness enhancements in the ring and correlated noise patches could also result in apparent, yet unphysical, offsets. To further probe this potential offset, we adopted an independent method to fit the outer ring morphology that does not take into account the absolute pixel brightnesses. We first determined the location of the central source by fitting a Gaussian profile to the central core of the \ec\ 70\,\mic\ image. We then considered the PSF-subtracted images of the outer ring, which we decomposed in 36 to 54 wedges. In each wedge, a radial profile was generated and fitted with either a Gaussian function or a polynomial of third to fifth order. The peak location in each wedge was recorded and an ellipse was fitted to these points (irrespective of the peak fluxes in each profile), providing estimates of the inclination, PA, semi-major axis and positional offset from the central source. We applied this method to both the original 70\,\mic\ image and to the (Lucy) deconvolved map. This second approach confirmed the existence and direction of the offset between the central star and the outer ring (readily evident in the deconvolved images), although its amplitude was estimated to be somewhat smaller, 0\farcs5$\pm$0\farcs2. While the methods lead to marginally consistent results, it appears that the amplitude of the offset is difficult to estimate precisely with the limited resolution of the PACS images. We further note that neither the PACS nor the SCUBA-2 images of the outer ring show evidence for a pericenter glow aligned with the direction of this possible offset, as could be expected \citep[e.g.,][]{wyatt99}. Higher resolution observations with ALMA are required to definitively assess the reality, direction and amplitude of any offset of the outer ring relative to the central star.


\subsection{Upper limits on scattered light brightness of the outer ring}
\label{subsec:hst}

For both the ACS and NICMOS datasets, we have visually inspected all PSF-subtracted images. Some structured residual artifacts, such as the modestly eccentric halo that is barely visible in the left panel of Figure\,\ref{fig:hst} and whose location unfortunately overlaps with the expected ring location, are often present in these images. These features are known artifacts, such as optical ghosts, imperfectly subtracted diffraction spiders and numerous radial spikes and concentric rings of positive and negative residuals, all of which are characteristic PSF subtraction errors \citep[e.g.,][]{clampin03, krist05}. They can be attributed to small changes in the {\it HST} optical path-length between the observations of \ec\ and the PSF star and/or slight color mismatches between these stars. The detailed appearances of these features are different in our multiple band-passes and vary substantially with each PSF template star; hence, they clearly are artificial in nature. In the absence of features that would be consistently present at multiple wavelengths and/or using multiple PSF stars, we conclude that there is no clear detection of scattered light at the {\it Herschel}-determined location of the outer ring of \ec. We note the presence of two faint point sources in the field-of-view of the ACS images located 13\arcsec\ and 13\farcs7 away from \ec\ at PA 256\degr and 83\degr, respectively. To the best of our knowledge, neither has been detected in the past so that it is not possible to estimate their proper motion and assess whether they are co-moving with \ec. Until follow-up observations are available, we assume that these are unrelated background sources.

\begin{figure}
\plotone{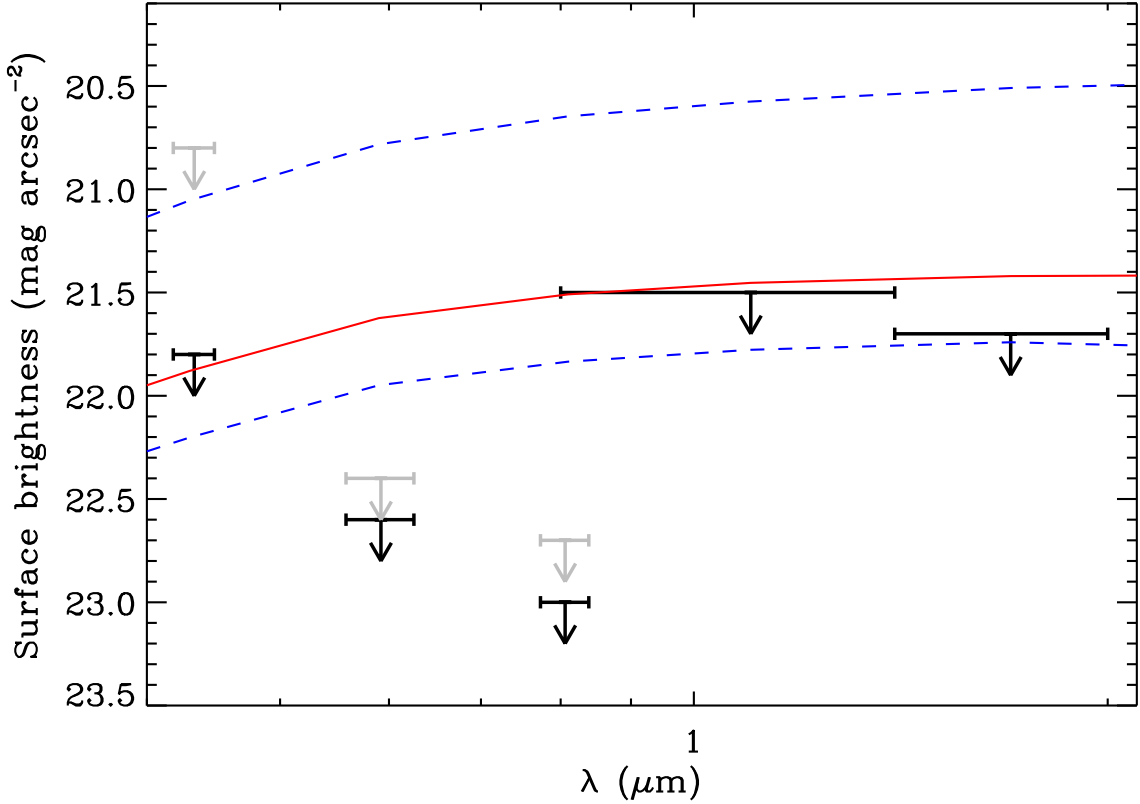}
\caption{Upper limits on the scattered light surface brightness of the \ec\ outer ring from our {\it HST} images estimated at the location of the ring ansae and in quadrature (black and gray arrows, respectively). Horizontal errorbars indicate the filter bandpasses. The solid red and dashed blue curves represent the predicted surface brightness (at the ansae and in quadrature, respectively) of the best fitting model derived in Section\,\ref{sec:model} on the basis of the system's SED and {\it Herschel} 70\,\mic\ image. Because of the system inclination and preference for forward scattering, the region of the disk that is inclined in front of the star is brighter than the back side by a factor of $\approx$3, which is illustrated by the difference between the two dashed blue curves. \label{fig:hst_lims}}
\end{figure}

We proceeded to place upper limits on the surface brightness of the ring. To do so, we created a grid of artificial 1\farcs0$\times$1\farcs0 constant surface brightness regions that we added at select locations along the expected location of the ring, namely along its minor and major axis. The surface brightnesses of these artificial sources were adjusted in an iterative manner until they became clearly visible in the image. For our ACS data, we consistently required a median signal to noise per pixel over the entire test region of at least 2 to claim a firm visual detection. Strong residual noise streaks in the PSF-subtracted NICMOS images led us to use a signal-to-noise per pixel threshold of 4. In the ACS datasets, increased residuals due to the 3\arcsec\ coronagraph spot are located almost exactly on the SE ansa of the ring, and we estimated our upper limit only on the NW ring. Similarly, because of the companion to HD\,105452, we estimated an upper limit along the minor axis only on the NE arc or the ring. In the NICMOS data, the limited field-of-view of the instrument forced us to estimate upper limits at the SE ansa only. The spectrum of \ec\ and the integration times used in our observing combine to yield upper limits on the scattered light brightness of the outer ring that are much deeper, by about 1\,mag/arcsec$^2$, at 0.6 and 0.8\,\mic\ than with the other filters (see Figure\,\ref{fig:hst_lims}). Specifically, the limiting surface brightness that we infer in these two filters at the location of the ring ansa are in the 22.6--23.0\,mag/arcsec$^2$ range. At the quadrature points in the ring (i.e., along the minor axis), the upper limits are lower by 0.2--0.3\,mag/arcsec$^2$ owing to the increased PSF subtraction residuals closer to the central star.


\section{Modeling}
\label{sec:model}


\subsection{Goals and Methodology}

The new {\it Herschel} images presented here provide the highest-resolution image of the outer debris ring in the \ec\ system to date. While SED modeling is generally fraught with ambiguities between dust grain size and distance to the central star, the high-quality spatial information contained in these images enables modeling of the dust ring that goes beyond the modified blackbody modeling presented above. Specifically, the 70\,\mic\ PACS image allows us to determine the radius of the dust ring precisely, thereby providing a tight constraint on the size distribution of the dust grains responsible for the far-infrared excess in this system. The inner ring can only be resolved with 100\,m-baseline mid-infrared interferometry \citep{smith09}, although the localization of the dust is not as precise for now. However, the mid-infrared spectrum of \ec\ strongly constrains the dust properties \citep{lisse12}. Therefore, it is now possible to perform a direct comparison of the dust properties between the two rings. Whether the two rings contain the same dust population has important implications for the physical relationship between them. 

To model the \ec\ system, we use MCFOST \citep{pinte06,pinte09} which treats the complete 3-dimensional dust radiative transfer problem assuming Mie theory. Briefly, the code evaluates the stellar radiation field and dust temperature throughout the disk, as well as the resulting spectral energy distribution, using a Monte Carlo approach, while synthetic images are computed using a ray tracing approach in a second step. The dust population is assumed to be in thermal equilibrium with the stellar radiation field. Because dust grains interact with light in a manner that depends on both wavelength and grain size, each grain size has a different equilibrium temperature. Large grains ($\gtrsim 10$\,\mic\ given our assumed dust composition) have temperatures that are close to the blackbody equilibrium temperature but smaller grains are substantially hotter.

Given the uncertainty surrounding a possible offset of the star relative to the outer ring, we assume that the ring is centered on the star, allowing us to adopt the 2-dimensional mode of MCFOST. Each dust belt is parametrized by its inner and outer radii, the surface density power law index (such that the surface density obeys $\Sigma(r) \propto r^{p}$), the total dust mass and the minimum and maximum grain size it contains. The grain size distribution is assumed to follow the canonical $N(a) \propto a^{-3.5}$ distribution \citep{dohnanyi69}, as supported by the analysis of the IRS spectrum of \ec\ \cite{lisse12}. There are therefore 12 parameters to explore in total; however, as in Section\,\ref{subsec:modbb}, the two dust rings are fit sequentially for a more efficient convergence. For both fits, the parameter space was explored first with a genetic algorithm that converged rapidly to a good family of solutions. We then ran an ensemble MCMC algorithm starting from one of these models to evaluate the uncertainties on, and correlation between, model parameters. After rejecting the "burning-in" phase of the chains, 50000 models are used to evaluate the posterior probability distributions. We used flat priors for the surface density power law index and the inner and outer radii of the belts, and log-flat priors for the minimum and maximum grain size (within the ranges $-2 \leq \log a_{min} \leq 2$ and $0 \leq \log a_{max} \leq 4$) and the total dust mass in each ring. The model with the lowest $\chi^2$ in each ensemble MCMC run (one for each belt) is considered the best fitting model; the associated parameters are listed in Table\,\ref{tab:modelpars}. As expected, the best model is comfortably within the confidence intervals derived from the associated ensemble MCMC run in both cases.

The dust composition of the warm dust ring was studied extensively by \cite{lisse12}. Although we have found that the absolute calibration of the IRS spectrum had to be re-scaled, the apparent morphology of the 10 and 20\,\mic\ silicate features are not severely affected. A thorough re-analysis of the dust composition is beyond the scope of the present analysis. Instead, we adopt a simplified dust composition that follows the spirit of the conclusions of \cite{lisse12}. Specifically, we assume that the dust grains can be described as a compact (non porous) mixture of astronomical silicates, amorphous carbon and water ice, with mass fractions of 70\%, 15\% and 15\%, respectively, resulting in a dust density of 2.9\,g/cm$^{3}$. Since we are interested in determining whether the two rings contain dust populations with a common origin, we assume this composition for both rings as a null hypothesis. The effective refraction index of the dust grains is computed assuming effective medium theory \citep{bruggeman35}.

Using the absorption and scattering cross-sections computed with Mie theory for a broad range of grain sizes, we self-consistently estimated the dust blowout size, defined as the grain size for which the ratio of the radiative and gravitational forces  is equal to $\beta_{rad/grav} = 0.5$, valid for particles initially on circular orbits. We find that $a_{blow} = 1.55\pm0.10$\,\mic, where the uncertainty stems from uncertainties on stellar properties ($\pm$0.03\,$R_\odot$ and $\pm$0.05\,$M_\odot$ for the stellar radius and mass, respectively). We do not a priori exclude that grains smaller than the blowout size are present in the system. However, such grains would be expelled from the system on a dynamical/orbital timescale, which is much shorter than the timescale on which they would be produced via collisions. 

One questionable feature of the dust model assumed here is the presence of a fraction of water ice in the dust grains. As pointed out by \cite{lisse12}, who first determined that this component had to be present to account for the IRS spectrum of \ec, the warm dust belt has a temperature that is substantially higher than the sublimation temperature of water ice. One possible explanation these authors offer is that the ice could be pure (i.e., separate grains made exclusively of ice). This is in apparent contradiction with our assumed dust model, in which the silicates, carbonaceous and water ice components are uniformly mixed in all dust grains. In that context, it would be more physically grounded to assume that water ice is present in the dust grains in the outer belt and that the corresponding ``identical" dust model for the inner belt consists of silicate-carbon grains with a porosity equal to the fraction of water ice present further out. To test the influence of this possible physical change, we have computed the SED of our best fit model for the inner belt with water ice replaced by void and found that it is statistically indistinguishable from that presented in Section\,\ref{subsec:model_inner} below. Therefore the conclusions reached in our analysis are not significantly sensitive to the presence or absence of water ice in the inner belt. 


\subsection{Modeling of the warm dust ring}
\label{subsec:model_inner}

The warm dust ring was fitted to the IRS spectrum limited to the 5.5 to 33\,\mic\ range to avoid the noisy edges of the spectrum and after re-sampling it log-uniformly to 30 independent points to smooth out the fine spectral features related to more complex dust components (e.g., crystalline silicates). We also included in the fit the 70\,\mic\ flux of the core component in the PACS image. Unlike the previous fit to the SED for the warm component, the flux for the 100\,\mic\ point was not enforced in
the $\chi^2_{warm}$, as the fit did not naturally exceed this limit. While the fitting process is primarily driven by this partial SED, we also took into account the fact that the inner ring has been spatially resolved by interferometric observations but is spatially unresolved in direct mid-infrared imaging \citep{smith09}. While the quality of that dataset was insufficient to precisely determine the geometry of the ring, it showed that most of the excess mid-infrared emission arose from the inner 3\,AU of the system. Thus, for each synthetic model, we computed the 11.5\,\mic\ image using the inclination derived for the outer ring and evaluated the fraction of the ring emission that lies within 3\,AU of the star. We added an arbitrarily large penalty term to the SED $\chi^2$ for those models where that fraction was less than 50\%, effectively rejecting them; otherwise the $\chi^2$ was not modified.

\begin{figure}
\plotone{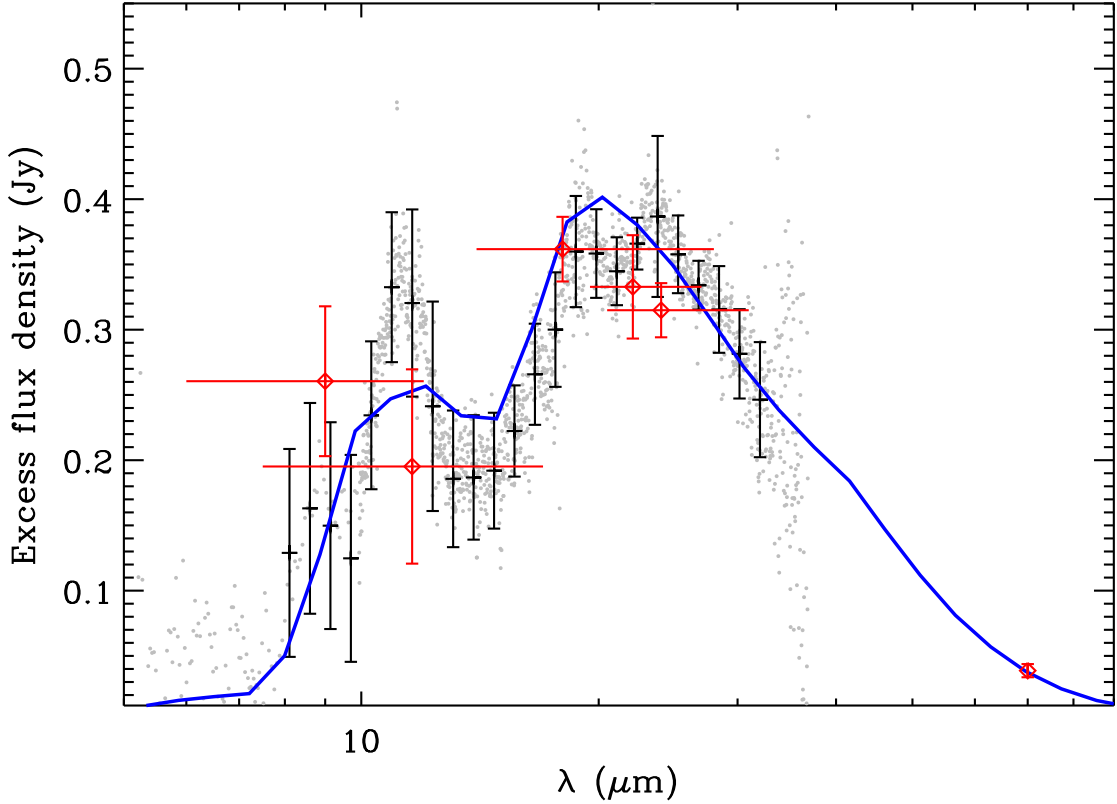}
\caption{Mid-infrared photometric and spectroscopic fluxes for \ec\ after subtraction of the stellar photosphere. The IRS spectrum is from \cite{lisse12} after scaling by a factor 0.86 to match broadband photometry from {\it AKARI}, {\it WISE} and MIPS (see Section\,\ref{subsec:obs_irs}) The black errorbars represent the re-binned IRS spectrum used for modeling in this work, with uncertainty estimated from the standard deviation within each bin. The red diamonds represent star-subtracted broadband photometry, including the 70\,\mic\ estimated for the `core component" of the PACS image. Vertical uncertainties compound a 2\% uncertainty on the estimated contribution of the stellar photosphere. The blue solid curve is the best fitting model of the inner ring from full radiative transfer. \label{fig:irs}}
\end{figure}

Although the fit is fairly good, the best fitting model (whose parameters are listed in Table\,\ref{tab:modelpars}) does not provide a perfect fit to the IRS spectrum (Figure\,\ref{fig:irs}). Most notably, the sharpness and strength of the 10-12\,\mic\ feature are under-predicted by the model. This is most likely because our compositional model and power law size distribution are too simplistic. We are nonetheless satisfied with the quality of the match and believe that this model captures the basic properties of the dust in the warm ring. Unfortunately, most of the model parameters are loosely constrained or affected by severe ambiguities, a typical conclusion for SED fitting alone. For instance, the outer radius of the inner ring is essentially unconstrained from the SED, as the surface density power law index can conspire to allow outer radii as large as 20\,AU, the largest outer radius we explored. Similarly, the inner radius posterior distribution is skewed towards radii larger than 2\,AU, consistent with the conclusions of \cite{smith09} and \cite{lisse12}, although much smaller radii are also statistically possible, as long as the minimum grain size is increased correspondingly. Indeed, there is a marked anti-correlation between $a_{min}$ and $R_{in}$. This is to be expected, as this prevents the hottest (smallest) dust grains from being exceedingly hot, and therefore generating a strong excess at shorter wavelengths. Finally, the total dust mass is strongly correlated with $a_{max}$ and thus only moderately constrained ($M_d \approx 10^{-6} - 10^{-5} M_\oplus$).

\begin{figure}
\plotone{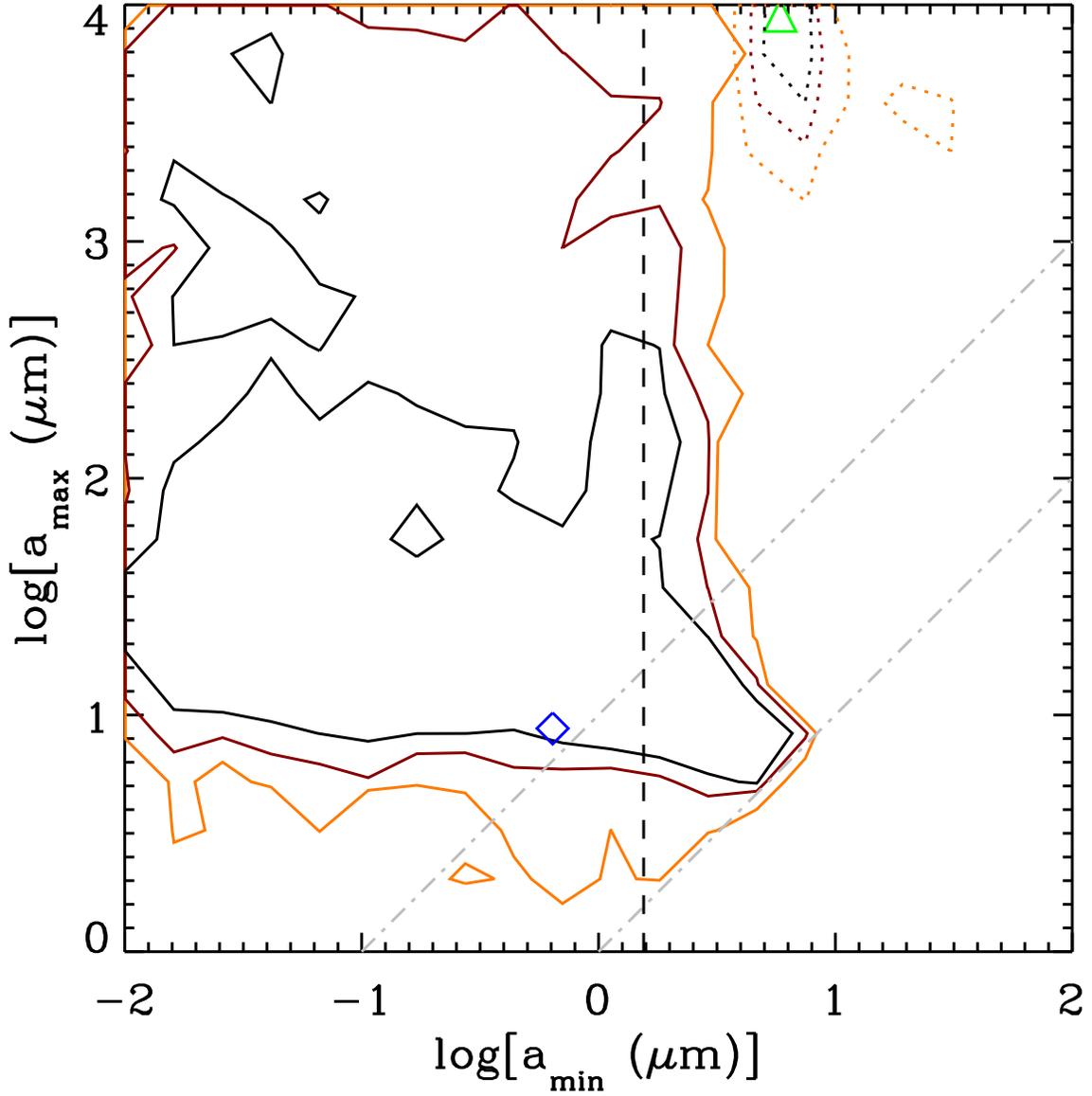}
\caption{Two dimensional posterior probability distributions for the minimum and maximum grain size in the inner (solid contours) and outer (dotted contours) belts of \ec. From dark to light, contours mark the 1, 2 and 3$\sigma$ contours for each dust population. The vertical black dashed line indicates the blow-out size while the slanted dotted gray lines mark the $a_{max} / a_{min} = 1$ and $a_{max} / a_{min} = 10$ limits. The blue diamond and green triangle mark the dust properties of the best fit model for the inner belt and outer belt, respectively. \label{fig:gsize_2d}}
\end{figure}

As shown in Figure\,\ref{fig:gsize_2d}, useful constraints are derived for both the minimum and maximum grain size, on the other hand. Most importantly, we find that the minimum grain size has to be small, with all values up to the blow-out size being equally probable (following our prior distribution). Intriguingly, some of the models that provide a good fit to the mid-infrared SED have narrow grain size distributions, with a maximum grain size only a few times larger than the minimum grain size. This subclass of models clusters around a minimum grain size of 2--5\,\mic\ and a maximum grain size of about 10--15\,\mic\ (reminiscent of the single grain size model considered in the modified blackbody fitting presented in Section\,\ref{subsec:modbb}). Such a narrow grain size distribution could be considered as physically improbable, so we also considered the subset of all models for which $a_{max} / a_{min} \geq 10$ as more likely to represent the physical state of the inner belt. For these models, the 2$\sigma$ upper limits on $a_{min}$ is 1.7\,\mic. Including the narrow grain size distributions, this upper limit only increases to 3.1\,\mic, a marginal difference. The maximum grain size has to be larger than a few microns, without notable preference, although we note a peak in the posterior distribution for $a_{max} \approx 10$\,\mic. Figure\,\ref{fig:gsize_2d} shows that the two parameters are only loosely correlated.

There is a ``ridge" of high posterior probability for models with $a_{min} \approx a_{blow}$ and $a_{max} \lesssim 1$\,mm. Such models are arguably the most plausible from a physical standpoint, but we emphasize that our modeling cannot exclude that much smaller dust grains are present in the inner belt with the data currently at hand. More precise mid-infrared interferometric observations of \ec\ would place a sharper constraint on the exact location of the warm dust belt, which in turn would translate into a better-defined minimum grain size in that belt. \cite{lisse12} concluded that the grain size distribution follows an $a^{-3.5}$ distribution down to about 1\,\mic, with the possibility of smaller grains being also present, albeit in lower numbers than predicted by extrapolating this power law. Our best fit model, with $a_{min} = 0.64$\,\mic, is in reasonable agreement with their conclusion despite the simplified dust composition we assumed here. From a physical standpoint, such a small minimum grain size is necessary to reproduce the strong mid-infrared silicate features. To summarize, we find that $a_{min} \lesssim a_{blow} \approx 1.5$\,\mic\ and $a_{max} \gtrsim 10$\,\mic\ in the inner belt.


\subsection{Modeling of the cold dust ring}
\label{subsec:model_outer}

As in Section\,\ref{subsec:geom}, we fit the outer dust ring by summing two separate $\chi^2$: one for the integrated SED at wavelengths $\lambda \geq 70$\,\mic, and the other for the 70\,\mic\ image using the same 30\arcsec$\times$30\arcsec\  field-of-view and pixel uncertainty as before. In this model, the warm ring is incorporated as a point source that is co-spatial with the star and whose flux is determined by the best fitting model from the previous section, while the outer ring is modeled as an inclined ring with the geometry parameters estimated in Section\,\ref{subsec:geom}. After convolution of the synthetic image with the {\it Herschel} PSF, the total flux in the map is normalized to unity, as was done for the observed image. This is done to focus the image fitting on the morphology of the image and not on its absolute flux. Indeed, a model over- or under-predicting the total flux for the system would already pay a penalty in the SED $\chi^2$. 

As illustrated in Figure\,\ref{fig:sed_mcfost}, the best fitting model yields an excellent fit to the PACS fluxes. However, the long wavelength slope of \ec\ is not very well reproduced. This is a consequence of our assumed dust composition and grain size distribution power law index since, even for very large values of $a_{max}$, the slope of the opacity function does not reach the Rayleigh-Jeans slope; the shallowest slope from such model has $\beta \approx 0.5$. Therefore the best-fitting model is a compromise, under-predicting the 850\,\mic\ flux but over-predicting the 250\,\mic\ one. This problem could be alleviated by changing the slope of the grain size distribution. Indeed, we find that using an $N(a) \propto a^{-3}$ size distribution leads to $\beta \approx 0$, as observed. However, to compensate for the reduced contribution of small dust grains to the total emission, fitting the whole SED of the outer ring then requires a physically improbable minimum grain size ($a_{min} \lesssim 0.02$\,\mic). While a better compromise might be obtained by considering the size distribution power law index as another free parameter, we consider that this is hardly justified considering that we only have six photometric datapoints beyond 70\,\mic\ (where the outer belt emission is detected) and already have six free parameters. Furthermore, a direct comparison of the dust properties in the two rings would be impossible without adding the same free parameter in the fit to the inner ring as well. In the following we only consider the model results associated with our nominal size distribution index.

\begin{figure}
\plotone{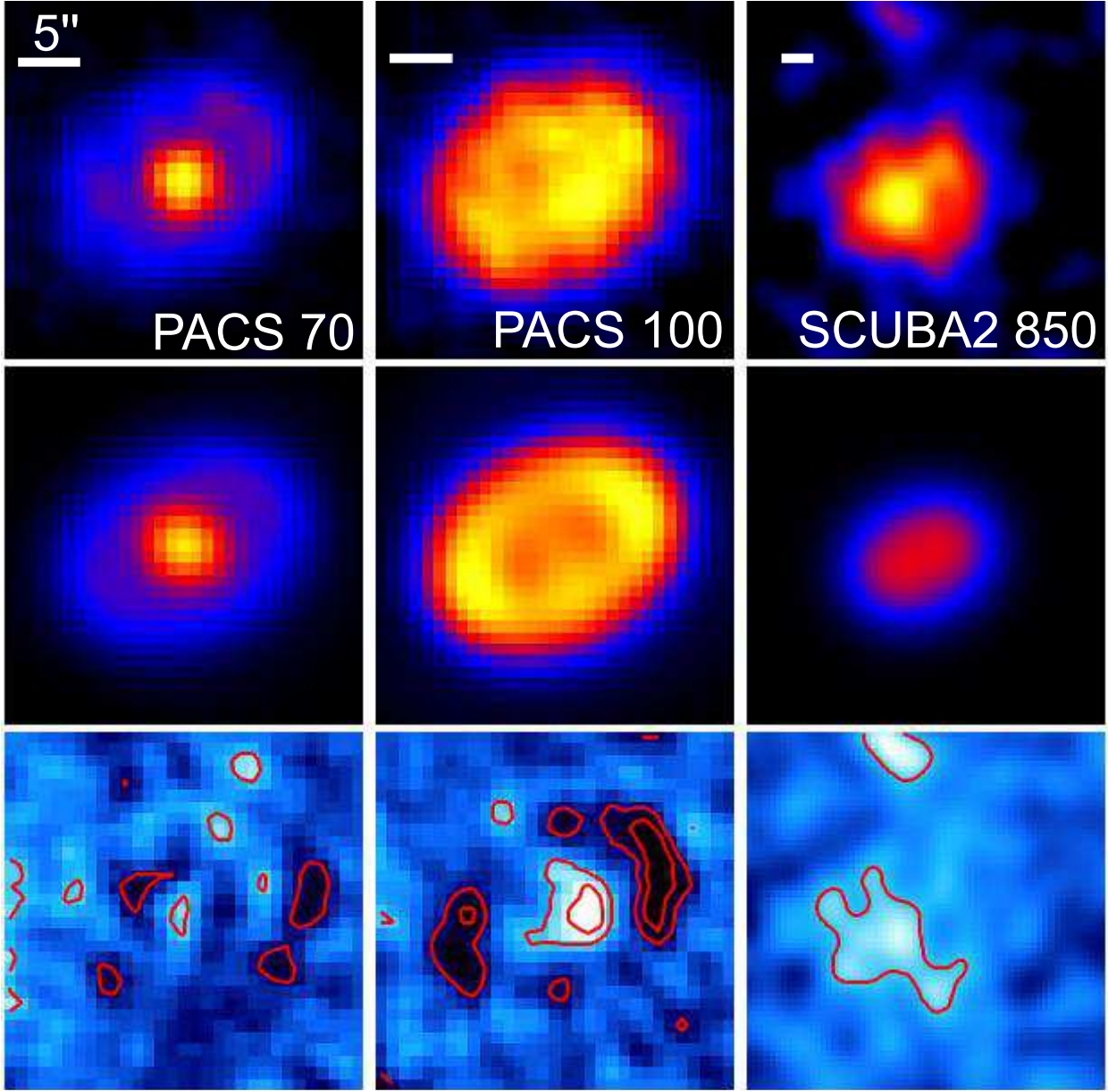}
\caption{{\it Top}: observed images of \ec\ at 70 and 100\,\mic\ from PACS and at 850\,\mic\ from SCUBA-2. The field of view of the panels is 30\arcsec\ for PACS and 60\arcsec\ for SCUBA-2, respectively. The images are shown on a linear stretch from 0 to 125\% of the peak intensity. {\it Middle}: synthetic images of the best-fitting model, including an inner point source and the cold outer ring, using the same relative color stretch as the top row. {\it Bottom}: residual maps after subtraction of the best fitting model, shown on a linear stretch from $-5\sigma$ to $5\sigma$, with contours at the $\pm3\sigma$ and $\pm5\sigma$. The only $5\sigma$ residuals are found at 100\,\mic, a dataset that was not included in the model fitting.\label{fig:resid}}
\end{figure}

The synthetic 70\,\mic\ image of the best-fitting model is an excellent fit to the observed image (see Fig\,\ref{fig:resid}). After subtraction of the flux-normalized model image, the residuals do not exceed $5\sigma$ anywhere in the map and are mostly random. The total $\chi^2$ for the image is 1315.3 over 961 pixels. The largest deviations between model and observations could be due to non-Gaussian (correlated) noise fluctuations or small-scale asymmetries within the ring. As an posteriori, independent check of the quality of the fit, we also computed residuals maps at 100 and 850\,\mic, which were not included in our fitting approach, but without applying any flux rescaling to allow for a more direct comparison. At 100\,\mic\, the same model does a reasonable job of reproducing the observed morphology. The strongest residual patch ($\approx 7 \sigma$ significance) is due to the fact that the brightest spot observed in the ring is not located on the ansae (as in the model image) but in quadrature instead. In addition to the same explanation as for the 70\,\mic\ image fitting, another shortcoming of our model could be that we underestimated the contribution of the inner belt at 100\,\mic. Indeed, our best-fitting model has a flux that is only about half of the $3\sigma$ upper limit we have derived in Section\,\ref{subsec:geom}. Finally, at 850\,\mic, our best fit model leads to marginally significant residuals ($\lesssim 2 \sigma$) in the SE ansa, a consequence of the fact that our best model produces a total flux that underestimates the observed one by 40\% or so. Still, since the significance of the increased brightness in that ansa is weak, we consider that our model provides a reasonable match to the system morphology at that wavelength.

\begin{figure}
\plotone{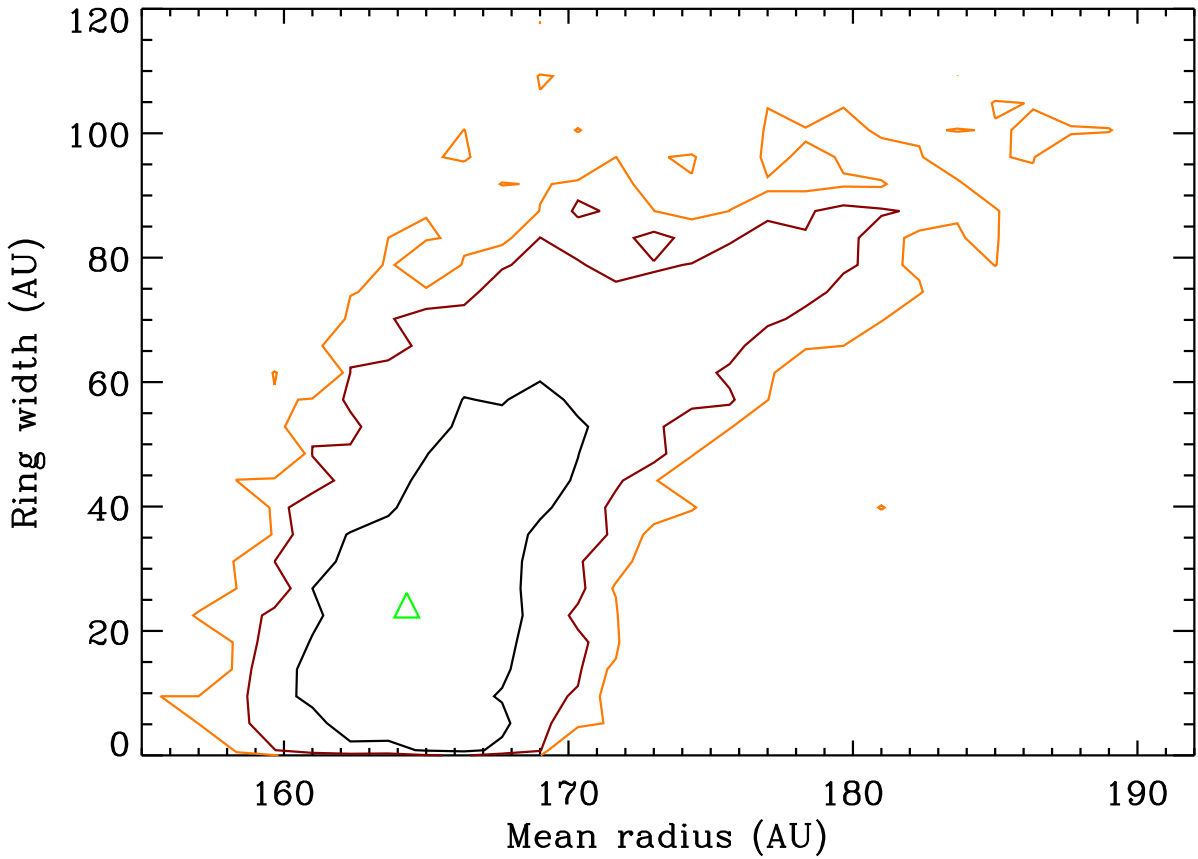}
\caption{Two dimensional posterior probability distributions for the mean radius and width of the cold dust ring. From dark to light, contours mark the 1, 2 and 3$\sigma$ contours for each dust population. The green triangle marks the properties of the best fit model. \label{fig:ring_bayesian}}
\end{figure}

The posterior distributions for most model parameters are well constrained, the only exception being the power law index of the surface density profile. Indeed, since the outer ring is not well resolved in our data, we can only place an upper limit on its width and the surface density profile within the ring remains undetermined. The 2$\sigma$ upper limits on the width of the outer ring is 75\,AU. The mean radius, defined as the mid-point between the inner and outer ring radii, is well constrained: $R_{mean} = 165.8^{+3.7}_{-2.9}$\,AU, with uncertainties indicating the 1$\sigma$ confidence interval. The two parameters are only weakly correlated (Figure\,\ref{fig:ring_bayesian}). These results are in excellent agreement with those derived from the purely geometrical models discussed in Section\,\ref{subsec:geom}, albeit with somewhat larger uncertainties. We can only place a lower limit on the maximum grain size, which is very close to the edge of our explored parameter space: $a_{max} \ge 3.5$\,mm at the 2$\sigma$ confidence level. The presence of much larger grains cannot be directly inferred with the data at hand since they do not extend beyond a wavelength of 850\,\mic. As a consequence, we can only place a lower limit on the total dust mass of 0.025\,$M_\oplus$ since that quantity is positively correlated with $a_{max}$.

The minimum grain size in the outer belt is very well constrained, $a_{min} = 6.4^{+0.7}_{-0.6}$\,\mic. This tight constraint results from the well determined dust temperature and distance of the dust from the star. With our dust composition, grains of size $a_{min}$ have an equilibrium temperature of 42\,K at the inner edge of the outer ring, in excellent agreement with the value derived from our modified blackbody fitting (see Section\,\ref{subsec:modbb}). On the other hand, in the context of our power law grain size distribution, we strongly reject the hypothesis that grains as small as the blowout size are present in the outer ring as those would reach a temperature of 59\,K and would yield a very poor fit to the spectral slope of the system between 70 and 160\,\mic, unless the grain size distribution is much flatter than we have assumed. Furthermore, the combination of minimum and maximum grain size we derive for the outer belt is inconsistent with those of the inner belt (see Figure\,\ref{fig:gsize_2d}). 

\begin{figure}
\plotone{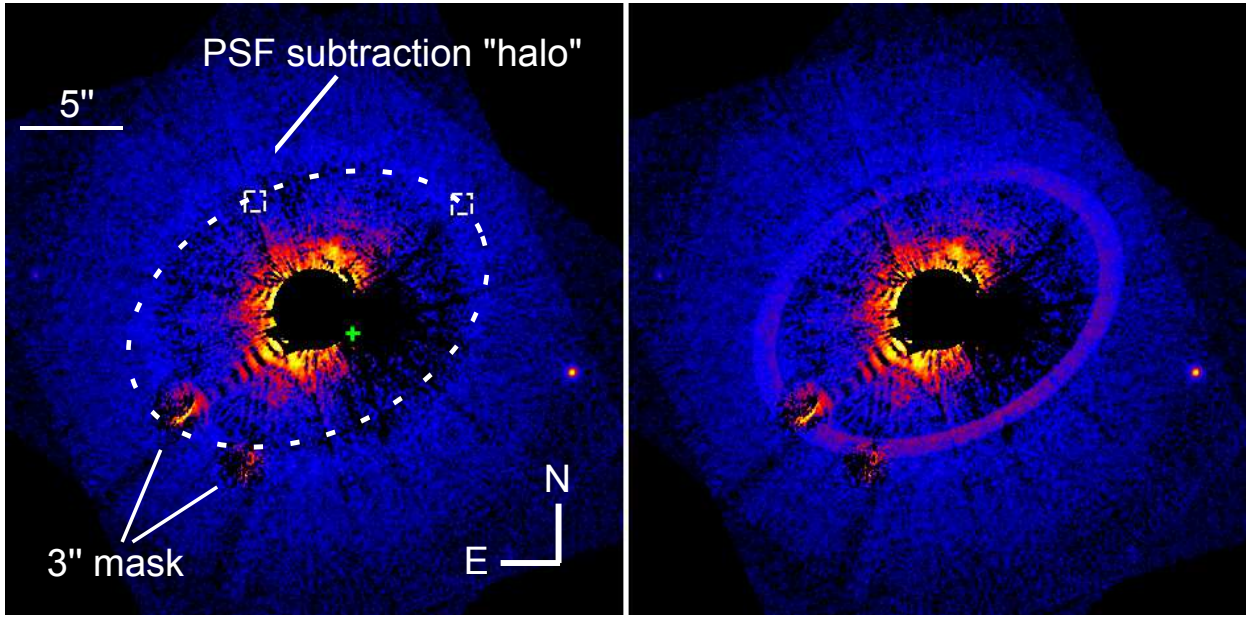}
\caption{(Left) Final, PSF-subtracted {\it HST}/ACS F606W image of \ec\ and median-smoothed using a 5$\times$5 running window. The field of view is 30\arcsec\ on a side. A central 2\arcsec-radius region has been masked out since it is dominated by PSF subtraction residuals. The location of the outer ring, as determined in the {\it Herschel} 70\,\mic\ image, is indicated by the dashed ellipse. The dashed white boxes indicate the regions used to estimate upper limits on the outer ring scattered light surface brightness at the ansa and in quadrature (to the NW and NE of the star, respectively). The green cross marks the location of the previously unnoticed faint background star in the vicinity of HD\,105452, the PSF star, which is also responsible for the systematic over-subtraction to the West of \ec. Strong residuals due to the 3\arcsec\ coronagraphic mask of ACS and a faint residual PSF subtraction halo are also indicated. (Right) Same image after injection of the best fit full radiative transfer model for the outer ring. We arbitrarily assumed that the region of the disk that is located in front of the star is located to the SW. That region thus appears brighter in scattered light because of the propensity of dust grains for forward scattering. With the model parameters listed in Table\,\ref{tab:modelpars}, the outer ring would have been detected in our {\it HST}/ACS observations. \label{fig:hst}}
\end{figure}

As an a posteriori test of the validity of this model, we have computed visible and near-infrared synthetic images for our best fit model. Because dust scattering occurs preferentially in the forward direction, the brightest regions in the belt are located on the near side of the ring, along the semi-minor axis. The easiest regions to detect, however, are located in the ansae along the semi-major axis owing to the larger distance from the central star, where a higher contrast can be achieved. From our best fit model, we find peak surface brightnesses in the ring ansae of about 21.5\,mag/arcsec$^2$, with relatively little chromaticity as expected for such large grains (approximately gray scattering). While this is essentially consistent with the upper limits derived form our {\it HST} images at 0.4, 1.1 and 1.6\,\mic, the predicted surface brightness for our best model at 0.6 and 0.8\,\mic\ violate the empirical upper limits by 1--1.5\,mag/arcsec$^2$ (see Figures\,\ref{fig:hst} and \ref{fig:hst_lims}). Similarly, the surface brightness in the "quadrature" regions (along the disk minor axis) exceeds the empirical upper limits. It must be noted that we do now know which side of the ring is actually tilted towards us but could only derive a meaningful upper limit only on one side of the star. However it is clear from Fig.\,\ref{fig:hst} that a disk at the predicted level would have been detected independently of this ambiguity.

The tension between the model surface brightness and the empirical upper limits can be partially alleviated by considering that our best fit model has a width of only 20\,AU. If the belt is three times as wide, which is acceptable at the 1$\sigma$ level, the resulting model surface brightnesses would be $\approx$1.2\,mag/arcsec$^2$ fainter, leaving a much more modest inconsistency to be explained. Furthermore, our modeling hinges exclusively on the absorption and emission dust properties, not on their scattering properties. It is plausible that relatively minor changes in dust composition, porosity, size distribution and/or dust grain shape could alter the dust scattering properties much more than their absorption/emission properties, most importantly their albedo \citep[e.g.][]{krist10}. Indeed, all other things being equal, the scattering surface brightness is proportional to the albedo. The average albedo, integrated over all scattering angles, of the dust population for our best model is about 0.55 across the visible and near-infrared ranges. Thus a dust population with an albedo $\lesssim 0.2$, as observed in a number of debris disks with scattered light detection \citep[][and references therein]{meyer07}, would not have been detected in our {\it HST} images. However, modifying the dust albedo without affecting the scattering phase function is impossible, which precludes using the {\it HST} non-detection to place a firm upper limit on the dust albedo. Broadly speaking, though, the scattered light surface brightness decreases rapidly as $a_{min}$ increases in the regime where $a_{min} \gtrsim \lambda$. This provides additional circumstantial evidence that micron-sized grains are absent in the outer belt. In the absence of a scattered light detection of the belt, we do not attempt to include the {\it HST} upper limits in our modeling, but we note that this non-detection favors a somewhat broader belt and/or inefficient scatterers.


\section{Discussion}
\label{sec:discus}


\subsection{The dust population in the outer belt}

The {\it Herschel} images of \ec\ presented here provide the highest resolution view of the cold dust ring in the system. Overall, the geometry of the \ec\ outer ring is strongly reminiscent of the one surrounding the A-type star Fomalhaut \citep{kalas05, kalas13}, including a possible offset between the geometrical center of the belt and the location of the central star. The latter object is much closer from the Sun than \ec, which has allowed mapping on smaller spatial scales with {\it Herschel} and ALMA \citep{acke12, boley12}. A detailed comparison of these two systems awaits future ALMA observations of \ec\ at higher resolution to cleanly separate the two rings, resolve the ring width, confirm the marginal offset and asymmetries, and place more stringent constraints on the presence of dust at intermediate radii. Nonetheless, as has been proposed for other two-ring systems \citep[e.g.,][]{su13}, it is plausible that yet undetected giant planets are responsible for the overall architecture of the \ec\ system, although the information at hand is insufficient to place any useful constraint at the moment \citep{lagrange09b}.

Building on the tight constraints on the spatial extent of both belts, the modeling presented here has enabled us to place stringent constraints on the dust properties in each dust ring. Within the context of our assumed dust model, we find that the outer belt is characterized by a minimum grain size that is several times larger than that found in the inner belt and than the blowout size. The apparent lack of grains with size in the range 1--4\,$a_{blow}$ in the outer ring is intriguing, since such grains should not be expelled from the system through radiation forces. We caution that some of our quantitative conclusions regarding grain sizes are dependent on the power law index of the grain size distribution and, more broadly, on the assumption of a pure power law size distribution, as well as on the assumed dust composition. Changing either of these parameters would quantitatively affect both the blowout size and the minimum and/or maximum grain sizes derived from our model fitting. For instance, assuming a flatter size distribution power law exponent increases the relative importance of large grains and would lead to a smaller minimum grain size. Indeed, a coarse parameter space exploration using a $N(a) \propto a^{-3}$ size distribution in the outer belt suggests that $a_{min} \leq 0.5$\,\mic, i.e., lower than the blow-out size for the system. The maximum grain size is also reduced, but only down to about 0.6\,mm because of the requirement to match the shallow sub-millimeter slope. 

Furthermore, even in an idealized steady-state collisional cascade, a perfect $N(a) \propto a^{-3.5}$ cannot extend all the way to the blow-out size. Indeed, the absence of smaller grains introduces an asymmetry in the collisional equilibrium that results in an over-density of grains somewhat larger than the blow-out size, as well as undulations about the nominal steady state power law \citep[e.g.,][]{thebault03}. If one fits a simple power law to such a function, it is possible that one would derive an apparent minimum grain size that is larger than the blow-out size. Thus, the assumption of a single power law size distribution uniformly populating a radially extended belt is not very physical. Instead, the interplay of collisions and radiative forces naturally results in a dust population characterized by a spatially-dependent grain size distribution that differs from a pure power law at any location \citep[e.g.,][]{thebault07}. However, the factor of 4 gap between $a_{blow}$ and $a_{min}$ appears quite wide. For instance, \cite{thebault07} find that the first peak in the size distribution is found at about 1.5 times the blow-out size. Therefore, it appears that another physical mechanism is responsible for the removal of grains up to a few times larger than the blow-out size in the outer belt on a timescale that is shorter than the collisional timescale. This situation is reminiscent of other debris disks \citep[e.g.,][]{lohne12} and is therefore not unique to \ec. 

Possible scenarios to account for this lack of grains above the blow-out size include Poynting-Robertson or stellar wind drag \citep{wyatt11}, photosputtering of icy grains \citep{lohne12} or a high eccentricity for the population of parent bodies \citep{wyatt10}. Instead of the removal of a select range of grain sizes, another possible interpretation is that the overall grain size distribution is shallower than we have assumed. Indeed, the model using $N(a) \propto a^{-3}$ discussed above produced a better overall fit to the SED of the system (except for the systematically low 250\,\mic\ SPIRE flux), as it also matched the observed shallow long wavelength slope while having a minimum grain size that is smaller than the blow-out size. However, this shallower size distribution also deviates from expectations of a collisional cascade, which is usually considered to be steeper, rather than shallower, than 3.5 in this size range, thus requiring additional physics. In fact, compared to the removal of only grains in the 1--5\,\mic\ range, accounting for such a shallow distribution necessitates affecting the dust population at all sizes up to millimeter sizes, which may be even more difficult to explain. Nonetheless, it is worth pointing out the inherent contradiction between the shallow millimeter slope, typical of large grains emitting like blackbodies, and the fact that the dust temperature in the outer belt is significantly higher than expected for blackbody equilibrium, which suggests the presence of copious amounts of relatively small grains. Accounting for both properties at once is a serious challenge and may require a size distribution whose power law slope changes at some intermediate size (in the 10--1000\,\mic\ range), with smaller grains characterized by a steeper distribution. Exploring this possibility is beyond the scope of our modeling.


\subsection{On the origin of the inner belt}

Before addressing the dust content of inner belt in more detail, it is worth noting that the {\it IRAS} 25 and 60\,\mic\ measurements that first revealed the presence of dust around \ec\ are fully consistent with the fluxes estimated with {\it Spitzer} and {\it Herschel} some 20 to 30\,yr later. This suggests that there is little to no variability on timescales of a few decades in the amount of emission from the inner belt, which dominates the excess emission at 25\,\mic\ and contributes non-negligibly to the 60--70\,\mic\ emission. Therefore, this apparent lack of variability allows us to place a lower limit on the lifetime of the inner belt that is much longer than its dynamical (orbital) timescale. Variations of excess emission on such short timescales have been observed in some debris disks, albeit not systematically \citep{beichman11, melis12, meng12}. 

Given the likely departure from pure power law dust populations, our conclusion that $a_{min} \leq a_{blow}$ in the inner belt indicates that grains smaller than the blow-out size are likely present in the inner ring and thus it appears that this belt cannot be described as a steady-state collisional cascade. Together with the copious amount of dust present in this ring \citep{wyatt07b}, this reinforces the argument that its origin is either a recent violent collision between very large bodies that released huge amounts of dust \citep{lisse12}, or that it is constantly replenished through another mechanism than the typical collisional cascade in a quiescent dusty belt \citep{bonsor11, bonsor13}.

Fundamentally the persistence of grains below the blow-out limit (i.e., those for which $\beta_{rad/grav} > 0.5$) for timescales longer than the dynamical timescales on which they would be lost is a serious issue which applies to most of the hot debris disks \citep[e.g.,][]{lisse08, lisse09, melis12}, and has three possible families of solutions: (i) the grains are being produced on a timescale that is shorter than the dynamical timescale, (ii) the radiation pressure acting on the dust grains is somehow reduced relative to naive expectations, or (iii) there is another force acting on the dust that is stronger than radiation pressure.

The first situation might occur in the period immediately following a giant impact, if that impact resulted in a large quantity of mm-sized vapor condensates. The collisional lifetime of those particles is shorter than one might expect if they were distributed in an axisymmetric disk, as all particles pass through the point at which the impact occurred. \cite{jackson12} found that the small dust resulting from their destruction could remain bright for many orbital periods, probably consistent with the empirical lower limit on the lifetime of the inner belt.

The second scenario might occur if the disk was optically thick in the radial direction, which would also require it to be vertically thin. The fractional luminosity of the inner disk implies that particle inclinations would need to be below 0\fdg01 in this scenario. Such a bright dynamically cool particle distribution could not persist so close to the star over a Gyr timescale \citep{krivov13}, and it is unclear that transient dust production mechanisms could produce copious dynamically cold dust. For instance, particles released in collisions would have random velocities of order the escape velocity of the parent body and so the required inclinations imply a parent body smaller than about 3\,km. This corresponds to a mass on the order of $6 \cdot 10^{-11}\, M_\oplus$, which is several orders of magnitude lower than the dust mass we have inferred for the inner belt.

The third possibility might arise, for instance, if the dust were orbiting within a gas disk that was dense enough for the dust to be coupled with the gas. For this scenario to take place, the dust grains with size $a_{blow}$ would have to have a stopping time that is equal to (or shorter than) the orbital timescale. We estimated the stopping time using Eq.\,16 in \cite{alexander07}, using a 3\,g/cm$^3$ grain density and assuming that the inner belt extends from 1 to 3\,AU from the central star with a flat surface density distribution. We find that that a total gas mass on the order of $2 \cdot 10^{-4}\,M_\oplus$ is required, which is two orders of magnitude larger than the dust mass we have inferred. Since debris disk gas has only been detected in young systems \citep[$\lesssim 30$\,Myr,][and references therein]{kospal13}, it would be highly unusual to find such a gas-rich debris disk in a system as old as \ec. There is no evidence to date for gas in this system, although it must be acknowledged that the current observational constraints are not very strict. 

The first scenario above appears to be the most likely one at this stage \citep[see also][]{lisse12}. However, we also have to consider the possibility that grains with $\beta_{rad/grav} > 0.5$ really are absent from the inner belt, despite the evidence presented above. This alternative explanation is indeed allowed by our modeling, since we find that a size distribution in which all of the particles are just above the blow-out limit can also fit the observations. We note, however, that a more detailed modeling of the IRS spectrum of \ec\ suggests that the grain size does extend at least up to 100\,\mic\ \citep{lisse12}, which seems to exclude this family of solutions. The physical origin of such a narrow particle distribution is less clear. However, the dearth of 1--4\,$a_{blow}$ grains in the outer belt, and the preponderance of such grains in the inner belt suggests a solution in which the smallest particles from the outer belt are transported into the inner regions. The most obvious mechanism for this is a drag force, and while P-R drag operates too slowly compared with collisions in the outer belt \citep{wyatt05b}, this could be increased say by stellar wind drag \citep{chen06, reidemeister11}. Drag forces alone could not explain the density enhancement in the inner regions, since it would result in a flat surface density distribution, meaning that another mechanism would need to be invoked to halt the dust once it had arrived in the inner regions. It is tempting to suggest that such a mechanism is resonant trapping by a giant planet, for which the limits for now are only moderately constraining \citep{lagrange09b}. However, not only is trapping unlikely for particles moving rapidly by drag forces \citep[e.g.,][]{mustill11}, but also particles inevitably fall out of resonance on timescales that are typically only an order of magnitude longer than the drag timescale, meaning that only a modest density enhancement is possible.

A more direct link between the outer and inner belts was suggested by \cite{wyatt10}, in that they are in fact derived from parent planetesimals that are on orbits with extreme eccentricities that take them all the way from the inner regions to the outer regions. This encompasses solution (i) above, by causing an enhanced collision rate in the inner belt with which to rapidly replenish small grains below the blow-out limit. While that model was consistent with all of the data available at the time, and has also been shown to be consistent with the $\beta$ Leo disk \citep{churcher11}, implicit in this model is that there should be thermal emission from the intermediate (3--100\,AU) region. The exact amount of such emission is model dependent, but a reasonable approximation is that the outer ring would be expected to emit about one third of the total 70\,\mic\ flux in the system, with the rest of the flux split roughly equally between the innermost and intermediate regions. Thus this model is ruled out by our 70\,\mic\ observations that set strong upper limits on the flux in the intermediate region. This does not completely rule out the possibility of a direct link between the inner and outer regions, however, since those planetesimals could be passed in through interactions with planets as are comets in the Solar System \citep[e.g.,][]{levison97}. This could either be a steady state situation for specific planetary system parameters \citep{bonsor13}, or a one off event analogous to the Late Heavy Bombardment in the Solar System \citep{booth09}. However, in that case one would again need to invoke solution (i) above, to enhance the production rate of blow-out grains, perhaps through the sublimation and disintegration of comet-like bodies.


\section{Conclusion}

As part of the DEBRIS survey, we have obtained new {\it Herschel} 70--500\,\mic\ images of the \ec\ debris disk system, as well as a new ground-based 850\,\mic\ map with SCUBA-2. The PACS 70\,\mic\ image is the highest resolution image to date of the thermal emission from the system and it allows us to disentangle the emission from the warm and cold dust belts, as well as to precisely constrain the geometry of the latter ring. Specifically, we determine the inclination and PA (47\degr\ and 116\degr, respectively) of the ring to within 1--2\degr\ and the midpoint radius of the ring ($\approx$165\,AU) with a precision of about 2\%. This radius is about twice as large as would be expected for dust grains emitting like blackbodies, indicating that significant amounts of relatively small grains are present in this belt. This is in contrast with the finding that the sub-millimeter spectral index is consistent with the Rayleigh-Jeans tail, indicative of large dust grains (mm-sized and larger). We can only place an upper limit on the ring width, namely 75\,AU at the 2$\sigma$ confidence level, because of the still limited resolution of {\it Herschel}. In addition, we find marginal evidence for azimuthal variations in the outer ring, as well as a possible offset of the center of the outer dust belt relative to the central star. While future higher resolution observations with ALMA are necessary to confirm these features, we note that they could be revealing the presence of unseen planets in the system. We also present deep {\it HST} visible and near-infrared coronagraphic images of the \ec\ system. The outer ring is not detected in these images and we place upper limit on its scattered light surface brightness at wavelengths ranging from 0.4 to 1.7\,\mic.

To interpret these observations, we construct full radiative transfer models of the system's SED and 70\,\mic\ image in an effort to constrain the dust properties in both belts. Assuming a simple power law grain size distribution, we find that the minimum grain size in the outer belt is about four times larger than the blow-out size, a conclusion that is also supported by the non-detection of scattered light in our {\it HST} images. While the gap between this minimum grain size and the blow-out size could be bridged if the grain size distribution was shallower than predicted by collisional cascade models, it is evident that some mechanism must act to remove from the outer belt grains that should not be expelled by radiation pressure alone. The available data do not allow us to determine which mechanism is most plausible. On the other hand, the dust in the inner belt has to contain substantial amounts of grains at the blow-out size and, quite possibly, even below this limit. Together with the very strong emission from the inner belt, this difference in dust properties between the two dust belts suggests that the warm dust cannot be explained by mere transport of grains from the outer belt in (e.g., via a production from eccentric planetesimals), but rather support a rare and more violent recent event. However, the time baseline between the first {\it IRAS} detection of the infrared excess in the system and the newest {\it Herschel} observations presented here shows that the dust in the inner belt must survive for several decades or be continuously replenished.


\acknowledgments We are grateful to Tushar Mittal, Christine Chen and Karl Stapelfeldt for discussions regarding various aspects of the data analyzed here, and to Angelo Ricarte and Noel Moldvai for their contribution to the modified blackbody model used in this work. We thank the DEBRIS team for many and varied fruitful discussions throughout the duration of this project. In particular, we are grateful to Paul Harvey for his review of a draft of this manuscript. Comments from an anonymous referee also helped improved this manuscript. This work was supported in part by NASA through a contract (No. 1353184, PI: H. M. Butner) issued by the Jet Propulsion Laboratory, California Institute of Technology under contract with NASA. We acknowledge the Service Commun de Calcul Intensif de l'Observatoire de Grenoble (SCCI) for computations on the super-computer funded by ANR (contracts ANR-07-BLAN-0221, ANR-2010-JCJC-0504-01 and ANR-2010-JCJC-0501-01) and  the European Commission's 7$^\mathrm{th}$ Framework Program (contract PERG06-GA-2009-256513). MW and GK are grateful for support from the European Union through ERC grant number 279973. CL acknowledges support from grants NASA NNX11AB21G and NSF AAG-NNX09AU31G in working on this project. PK acknowledges support from NASA NNX11AD21G, NSF AST-0909188 and JPL/NASA award NMO711043. MB acknowledges support from an NSERC Discovery Accelerator Supplement. Data presented in this paper were obtained with {\it Herschel}, an ESA space observatory with science instruments provided by European-led Principal Investigator consortia and with important participation from NASA. Additional data presented in this work were obtained at JCMT, which is operated by the Joint Astronomy Centre on behalf of the Science and Technology Facilities Council of the UK, the Netherlands Organisation for Scientific Research and the National Research Council of Canada. Additional funds for the construction of SCUBA-2 were provided by the Canada Foundation for Innovation. This publication makes use of data products from the Wide-field Infrared Survey Explorer, which is a joint project of the University of California, Los Angeles, and the Jet Propulsion Laboratory/California Institute of Technology, funded by the National Aeronautics and Space Administration. This research has made use of the SIMBAD database, operated at CDS, Strasbourg, France, and of the NASA/IPAC Infrared Science Archive, which is operated by the Jet Propulsion Laboratory, California Institute of Technology, under contract with the National Aeronautics and Space Administration.

{\it Facilities:} \facility{Herschel}, \facility{JCMT}, \facility{HST}.

\bibliography{gduchene}

\begin{deluxetable}{ccccc}
\tablecaption{Details of {\it Herschel} observations of \ec.\label{tab:log}} 
\tablewidth{0pt} 
\tablehead{
  \colhead{Instrument} & \colhead{Obs. ID} & \colhead{Obs. Date} &
  \colhead{$\lambda$ (\mic)} & \colhead{Duration (s)} }
\startdata
PACS & 1342222622--3 & 06/15/2011 & 70, 160 & 4$\times$445 \\
PACS & 1342234385--6 & 12/15/2011 & 100, 160 & 4$\times$445 \\
SPIRE & 1342212411 & 01/09/2011 & 250, 350, 500 & 721 \\
\enddata
\end{deluxetable}
\clearpage

\begin{deluxetable}{cccccc}
\tablecaption{Details of {\it HST} observations of \ec.\label{tab:hst}} 
\tablewidth{0pt} 
\tablehead{
  \colhead{Instrument} & \colhead{Obs. Date} & \colhead{Target} & \colhead{Sp.\,T.} &
  \colhead{Filter} & \colhead{Duration (s)} }
\startdata
\cutinhead{Observations taken as part of program 10244}
ACS/HRC & 2005/01/30 & \ec & F2V & F435W & 2$\times$2300 \\
 & & & & F606W & 2$\times$2275 \\
 & & & & F814W & 2$\times$2250 \\
 & 2005/01/29 & HD\,105452 & F1V & F435W & 2600 \\
 & & & & F606W & 2300 \\
 & & & & F814W & 2295 \\
NICMOS-2 & 2005/07/10 & \ec\ & F2V & F110W & 512 \\
 & & & & F160W & 512 \\
 & & & & F171M & 416 \\
 & 2005/08/15 & HD\,132052 & F0V & F110W & 416 \\
 & & & & F160W & 416 \\
 & & & & F171M & 416 \\
\cutinhead{Archival data of PSF template stars}
ACS/HRC & 2004/09/06 & HD\,27290\tablenotemark{a} & F1V & F606W & 2460 \\
 & 2006/02/25 & HD\,142860\tablenotemark{b} & F6IV & F606W & 900 \\
 & 2006/08/16 & HD\,68456\tablenotemark{c} & F6Ve & F606W & 2100 \\
NICMOS-2 & 2004/10/24 & HD\,38207\tablenotemark{d} & F2V & F110W & 224 \\
 & 2004/10/31 & HD\,35841\tablenotemark{d} & F3V & F110W & 224 \\
 & 2004/10/22 & HIP\,22844\tablenotemark{e} & F5V & F160W & 192 \\
 & 2004/11/02 & HIP\,24947\tablenotemark{e} & F6V & F160W & 192 \\
 & 2004/10/31 & HIP\,1134\tablenotemark{e} & F7V & F160W & 192 \\
 & 1998/10/20 & HD\,84117\tablenotemark{f} & F8V & F171M & 144 \\
\enddata
\tablenotetext{a}{Data taken as part of program 9475 (PI P. Kalas).}
\tablenotetext{b}{Data taken as part of program 10599 (PI P. Kalas).}
\tablenotetext{c}{Data taken as part of program 10896 (PI P. Kalas).}
\tablenotetext{d}{Data taken as part of program 10177 (PI G. Schneider).}
\tablenotetext{e}{Data taken as part of program 10176 (PI I. Song).}
\tablenotetext{f}{Data taken as part of program 7835 (PI E. Rosenthal).}
\end{deluxetable}
\clearpage

\begin{deluxetable}{ccc|ccc|ccc}
\tablecaption{SED of \ec.\label{tab:sed}} 
\tablewidth{0pt} 
\tablehead{
  \colhead{$\lambda$ (\mic)} & \colhead{$F_\nu$ (Jy)} & \colhead{Ref.} &
  \colhead{$\lambda$ (\mic)} & \colhead{$F_\nu$ (Jy)} & \colhead{Ref.} &
  \colhead{$\lambda$ (\mic)} & \colhead{$F_\nu$ (Jy)} & \colhead{Ref.} }
\startdata
0.36\tablenotemark{\dagger}\tablenotemark{a} & 23.3$\pm$1.2 & 1 & 9\tablenotemark{\dagger} & 2.32$\pm$0.04 & 3 & 100 & 0.805$\pm$0.084 & 6 \\
0.44\tablenotemark{\dagger}\tablenotemark{a} & 53.2$\pm$2.7 & 1 & 11.6\tablenotemark{\dagger}\tablenotemark{b} & 1.46$\pm$0.07 & 4 & 100\tablenotemark{\dagger}\tablenotemark{e} & 0.252$\pm$0.016 & 8 \\
0.55\tablenotemark{\dagger}\tablenotemark{a} & 66.2$\pm$3.3 & 1 & 11.9 & 1.51$\pm$0.24 & 5 & 160\tablenotemark{\dagger} & 0.231$\pm$0.013 & 8 \\
0.70\tablenotemark{\dagger}\tablenotemark{a} & 76.4$\pm$3.8 & 1 & 12 & 1.63$\pm$0.05 & 6 & 250\tablenotemark{\dagger} & 0.100$\pm$0.010 & 8 \\
0.90\tablenotemark{\dagger}\tablenotemark{a} & 70.5$\pm$3.5 & 1 & 18\tablenotemark{\dagger} & 0.82$\pm$0.02 & 3 & 350\tablenotemark{\dagger}\tablenotemark{f} & $\leq$0.10 & 8 \\
1.2\tablenotemark{\dagger}\tablenotemark{a} & 53.5$\pm$2.7 & 2 & 22\tablenotemark{\dagger}\tablenotemark{c} & 0.68$\pm$0.04 & 4 & 450\tablenotemark{\dagger} & 0.058$\pm$0.010 & 5 \\
1.65\tablenotemark{\dagger}\tablenotemark{a} & 38.1$\pm$1.9 & 2 & 24\tablenotemark{\dagger} & 0.59$\pm$0.02 & 7 & 500\tablenotemark{\dagger}\tablenotemark{f} & $\leq$0.07 & 8 \\
2.2\tablenotemark{\dagger}\tablenotemark{a} & 25.2$\pm$1.3 & 2 & 25 & 0.59$\pm$0.03 & 6 & 850\tablenotemark{\dagger} & 0.0155$\pm$0.014 & 8 \\
3.6\tablenotemark{\dagger}\tablenotemark{a} & 11.4$\pm$0.6 & 2 & 60 & 0.263$\pm$0.041 & 6 & 850 & 0.0143$\pm$0.0018 & 5 \\
3.8\tablenotemark{\dagger}\tablenotemark{a} & 9.6$\pm$0.5 & 2 & 70 & 0.198$\pm$0.007 & 7 & 850\tablenotemark{g} & 0.0075$\pm$0.0012 & 9 \\
4.8\tablenotemark{\dagger}\tablenotemark{a} & 6.0$\pm$0.3 & 2 & 70\tablenotemark{\dagger}\tablenotemark{d} & 0.230$\pm$0.013 & 8 & & & \\
\enddata
\tablenotetext{\dagger}{Flux included in the final composite SED of \ec. Other entries in this table are set aside due to lower quality, confusion or because they were superseded by more recent observations.}
\tablenotetext{a}{5\% flux uncertainty assumed.}
\tablenotetext{b}{4.5\% flux uncertainty assumed \citep{jarrett11}.}
\tablenotetext{c}{5.7\% flux uncertainty assumed \citep{jarrett11}.}
\tablenotetext{d}{Integrated flux for the system. The outer ring contributes 160$\pm$15\,mJy while the star and inner ring combine to a flux density of 70$\pm$10\,mJy.}
\tablenotetext{e}{Integrated flux for the system. The star and inner ring are not firmly detected and we place an upper limit of 45\,mJy on their combined flux density.}
\tablenotetext{f}{Substantial confusion with nearby background galaxies.}
\tablenotetext{g}{SCUBA observations obtained in ``photometry mode" which underestimates the flux of extended sources.}
\tablerefs{ (1) \cite{johnson66}; (2) \cite{sylvester96}; (3) {\it AKARI} All-Sky Catalog \citep{ishihara10}; (4) {\it WISE} All-Sky Catalog \citep{wright10}; (5) \cite{wyatt05}; (6) {\it IRAS} Faint Source Catalog \citep{moshir92}; (7) \cite{beichman06}; (8) This work; (9) \cite{sheret04}.}
\end{deluxetable}
\clearpage

\begin{deluxetable}{ccccc}
\tablecaption{Photometry and astrometry of sources in the immediate vicinity of \ec.\label{tab:galaxies}} 
\tablewidth{0pt} 
\tablehead{
  \colhead{Source} & \colhead{$\Delta$RA (\arcsec)} & \colhead{$\Delta$Dec (\arcsec)} &
  \colhead{$F_\nu^{100\mu{\rm m}}$ (mJy)} & \colhead{$F_\nu^{160\mu{\rm m}}$ (mJy)} }
\startdata
B & -3.4$\pm$0.2 & +20.3$\pm$0.2 & 4.6$\pm$1.5 & 30$\pm$6 \\
C & +24.7$\pm$0.1 & -41.4$\pm$0.1 & 18$\pm$3 & 28$\pm$5 \\
D & -5.4$\pm$0.2 & +31.6$\pm$0.2 & 5.7$\pm$1.7 & 32$\pm$6 \\
\enddata
\tablecomments{All positions are measured relative to the centroid position of \ec, using a weighted average to combine the astrometric information from the 100 and 160\,\mic\ images. Uncertainties only reflect the signal-to-noise of each detection and does not include terms associated to the morphology of the sources and of \ec\ itself. Flux uncertainties quadratically compound signal-to-noise, aperture correction and absolute flux calibration terms. These sources are identified in Figure\,\ref{fig:images}.}
\end{deluxetable}
\clearpage

\begin{deluxetable}{ccccc}
\tablecaption{Modeling parameters for the dust belts.\label{tab:modelpars}} 
\tablewidth{0pt} 
\tablehead{
  \colhead{Parameter} & \multicolumn{2}{c}{Inner Belt} & \multicolumn{2}{c}{Outer Belt} \\
 & Range & Best-fit & Range & Best-fit}
\startdata
\cutinhead{Geometric modeling}
$i$ (\degr) & -- & -- & 46.8$_{-1.2}^{+1.5}$& 46.8 \\
PA (\degr) & -- & -- & 116.3$_{-1.4}^{+1.7}$& 116.3 \\
$R_{mid}$ (AU) & -- & -- & 164.2$\pm$2.5 & 163.6 \\
$\Delta R$ (AU) & -- & -- & $\leq 60$ & 9 \\
\cutinhead{Modified blackbody modeling}
$L_{\rm IR} / L_\star$ ($10^{-5}$) & 32.7$\pm$2.2 & 32.5 & 2.17$\pm$0.06 & 2.17 \\
$T$ (K) & 418$\pm$16 & 409 & 42.3$\pm$1.4 & 42.1 \\
$\log [\lambda_0$ (\mic)] & 1.7$^{+1.1}_{-0.0}$ & 1.8 & 0.5$^{+1.9}_{-0.6}$ & 0.12 \\
$\beta$ & $\geq 0.7$ & 1.5 & 0.21$\pm$0.07 & 0.22 \\
\cutinhead{Full radiative transfer}
$R_{mid}$ (AU) & 6.4$\pm$2.7 & 8.5 & 165.8$^{+3.7}_{-2.9}$ & 164.3 \\
$\Delta R$ (AU) & $\leq 17$ & 16 & $\leq 75$ & 22 \\
$p$ & -1.6$^{+1.7}_{-1.6}$ & 0.2 & -1.7$^{+1.9}_{-1.6}$ & -2.5 \\
$a_{min}$ (\mic) & $\leq 3.1$ & 0.64 & 6.4$^{+0.6}_{-0.7}$ & 5.9 \\
$a_{max}$ (\mic) & $\geq 8.0$ & 8.8 & $\geq 3500$ & 8980 \\
$M_d$ ($M_\oplus$) & $\geq 1.2 \cdot 10^{-6}$ & $4.3 \cdot 10^{-6}$ & $\geq0.016$ & 0.025 \\
\enddata
\tablecomments{For both dust belts, the first column indicates the 1$\sigma$ confidence interval for, or 2$\sigma$ upper limit on, each model parameter. The values listed under the ``best-fit" columns represent the sets of model parameter used to generate Figures\,\ref{fig:sed_mcfost}, \ref{fig:irs} and \ref{fig:resid}.}
\end{deluxetable}
\clearpage

\end{document}